\pdfoutput=1
\documentclass{article}
\usepackage{blindtext}
\usepackage{cite}
\usepackage{hyperref}
\hypersetup{%
	colorlinks,
	plainpages=false,
	pdfpagelabels,
	breaklinks,
	pdfview=Fit,
	bookmarksopen,
	bookmarksnumbered,
	linkcolor=black,
	anchorcolor=black,
	citecolor=black,
	filecolor=black,
	urlcolor=LPRred
}%
\usepackage{graphicx}
\usepackage{bm}
\usepackage{textgreek}
\usepackage{amsmath}

\usepackage{placeins}   

\begin{document}
	
	\title{Giant Gain Enhancement in Surface-Confined Resonant Stimulated Brillouin Scattering}
	
	\author{Nathan Dostart, Seunghwi Kim, Gaurav Bahl*\\
		\\
		Department of Mechanical Science and Engineering\\
		University of Illinois at Urbana-Champaign\\
		1206 W. Green St., Urbana, Illinois, USA 61801\\
		*Corresponding Author: bahl@illinois.edu}
	
	\maketitle
	
	\begin{abstract}
		The notion that Stimulated Brillouin Scattering (SBS) is primarily defined by bulk material properties has been overturned by recent work on nanoscale waveguides. It is now understood that boundary forces of radiation pressure and electrostriction appearing in such highly confined waveguides can make a significant contribution to the Brillouin gain. Here, this concept is extended to show that gain enhancement does not require nanoscale or subwavelength features, but generally appears where optical and acoustic fields are simultaneously confined near a free surface or material interface. This situation routinely occurs in whispering gallery resonators (WGRs), making gain enhancements much more accessible than previously thought. To illustrate this concept, the first full-vectorial analytic model for SBS in WGRs is developed, including optical boundary forces, and the SBS gain in common silica WGR geometries is computationally evaluated. These results predict that gains $10^4$ times greater than the predictions of scalar theory may appear in WGRs even in the 100 {\textmu}m size range. Further, trapezoidal cross-section microdisks can exhibit very large SBS gains approaching $10^2$ m\textsuperscript{-1}W\textsuperscript{-1}. With resonant amplification included, extreme gains on the order of $10^{12}$ m\textsuperscript{-1}W\textsuperscript{-1} may be realized, which is  $10^8$ times greater than the highest predicted gains in linear waveguide systems.
	\end{abstract}

\section{Introduction}
\label{sec:intro}
Stimulated Brillouin Scattering (SBS) is a fundamental acousto-optical nonlinearity \cite{Chiao1964,Shen1965,Yariv1965} that occurs in all states of matter and is a powerful tool for coupling traveling photons to traveling acoustic phonons. The process occurs through light scattering from photoelastic perturbations created by an acoustic wave in a medium, accompanied by simultaneous amplification (for Stokes scattering) or attenuation (for anti-Stokes scattering) of the acoustic wave through electrostrictive feedback. When sufficient optical power is pumped into the medium, the Stokes scattering occurs regeneratively and a Brillouin laser is generated at a fixed offset from the pump laser. In addition to being a method for generating \cite{Bahl2011}, suppressing \cite{Bahl2012Cool}, and interrogating \cite{Montrose1968,Lee1987,Scarcelli2008} acoustic modes, SBS has been employed effectively in a wide range of applications \cite{Kabakova2015} including ultra-narrow linewidth lasers \cite{Chiao1964}, quantum optics and electrodynamics experiments \cite{Schliesser2010,Milburn2011,Dong2012,Weis2010}, distributed temperature and pressure sensing \cite{Kurashima1990}, optical amplifiers \cite{Boyd2008} and modulators, photonic oscillators \cite{Tomes2009,Grudinin2009,Lee2012}, optical cooling \cite{Bahl2012Cool}, optical phase conjugation \cite{Zel1972}, “slow light” \cite{Kim15,Boyd2009}, and nonreciprocal optical devices \cite{Kim15,Kang2011,Poulton2014}. SBS is one of the highest gain material-level nonlinearities in optics \cite{Boyd2008} making it one of the first nonlinear effects to appear in high energy density systems such as high power fiber lasers and high quality factor resonators. The suppression of SBS is thus particularly important in laser systems as it represents a parasitic loss effect and prevents further power scaling of the source \cite{Smith1972,Dragic2012}.\par

We consider in this study the SBS gain, which describes the amplification or attenuation experienced by a small signal probe that is offset from the pump by the acoustic frequency. Textbook descriptions \cite{Boyd2008,Agarwal2013} state that the line-center SBS gain is defined by the electrostrictive constant of the material, the optical pump frequency, the refractive index, the speed of sound, and the phonon losses in the material. The role of geometry or boundaries is generally not considered. However, in recent theoretical \cite{Rakich2012} and experimental \cite{Shin2013,VanLaer2015} studies on devices with nanoscale dimensions, significant SBS gain enhancement has been deduced. This surprising result occurs due to very large radiation pressure and electrostriction forces at the device boundaries \cite{Qiu2013,Shin2013,Rakich2010,Rakich2012}. Here we propose that these boundary force enhancements will not only appear in sub-wavelength devices, but also whenever an optical field is confined near a material interface or free surface. Such surface-confined Brillouin scattering interactions are already known to occur in whispering gallery resonators (WGRs) as demonstrated in \cite{Tomes2009,Bahl2011,Bahl2012Cool,Bahl2013,Bahl2011Characterization,Kim15,Grudinin2009}. In microscale WGRs, light and traveling acoustic waves are both tightly confined near the resonator surface, even without the benefit of any sub-wavelength dimensions, thus very large SBS gains should be expected. A similar insight has also been developed recently \cite{Baker2014} in a study where the effects of boundary forces on axisymmetric stationary vibrational modes in WGRs were analyzed. Experimentally, the highest measured SBS gain of $4\times10^6$ m\textsuperscript{-1}W\textsuperscript{-1} has been observed in a microsphere WGR in a Forward SBS configuration (see Supplement Table S.1 in ref.~\cite{Kim15}), and it is important to understand how such large values of gain arise in such large systems.

In this paper, we demonstrate that the large optical forces generated at WGR boundaries can both enhance and attenuate the Brillouin gain by multiple orders-of-magnitude. We provide, for the first time, a full vectorial description of SBS gain in WGRs incorporating the effects of boundary forces including electrostriction and radiation pressure. We then use the formalism to computationally evaluate the SBS gain in silica WGRs of several commonly used geometries, across a range of size scales, while examining the individual contributions of the optical forces to the SBS gain.

\section{Theory of SBS in Linear Systems}
\label{sec:linear_theory}

SBS is a three-wave interaction between two optical fields and one traveling acoustic field in a dielectric medium \cite{Boyd2008,Shen1965}. Coupling between the waves is enabled primarily through photoelastic scattering and electrostriction. The two optical fields are selected such that they generate a spatiotemporal overlap matching the frequency and momentum associated with the acoustic wave, resulting in the phase matching requirement described below. This field overlap generates a spatially varying distribution of electrostrictive pressure. The gradient of this pressure generates optically-induced forces which can enhance (or even attenuate \cite{Bahl2012Cool}) the acoustic wave. The acoustic wave, in turn, creates a periodic refractive index perturbation in the medium through the photoelastic effect, through which light scattering between the optical fields can occur. The higher intensity, typically higher frequency optical field is called the ``pump'', while the scattered or probed optical field is called ``Stokes'' due to its lower frequency. SBS can also occur using optical signals of two different polarizations \cite{Kang2010,Kang2011}. When light scattering takes place in the forward direction, MHz - GHz frequency acoustic waves are involved, and this is called Forward SBS (F-SBS). Scattering in the backward direction is called Backward SBS (B-SBS), in which multi-GHz acoustic waves mediate the optical coupling. 

A key requirement in SBS is the energy and momentum phase matching. The mathematical relationship for Stokes light scattering is written as $\vec{k}_p-\vec{k}_s=\vec{K}$ and $\omega_p-\omega_s=\Omega$, where $\vec{k}_p$, $\vec{k}_s$, and $\vec{K}$ are the momentum vectors of the photon modes and the phonon mode respectively, while $\omega_p$, $\omega_s$, and $\Omega$ are the corresponding angular frequencies.

An additional requirement of SBS is that the optical forces be able to effectively drive the acoustic mode; this is analogous to having strong coupling of power from the pump beam into the Stokes and acoustic modes. The derivation in Section~\ref{sec:wgr_theory} is centered on calculating this coupling using the opto-acoustic overlap integral (see Eq.~\eqref{eq:2}) of the form $\left\langle F|U\right\rangle$. Good coupling occurs when the optical forces $F$ align well with the acoustic mode $U$, which is intuitively understood as efficient power transfer into the acoustic mode. An alternative, but entirely equivalent, method for determining the strength of this coupling is to use the optical coupling integral of the form $\left\langle E_1|\Delta \epsilon|E_2\right\rangle$.  The optical coupling integral describes how a perturbation $\Delta\epsilon$ to the dielectric permittivity (induced by the acoustic mode for SBS) facilitates power exchange between two optical modes $E_1$ and $E_2$. The equivalence of these two formulations was recently proved in the case of low loss and reversible optical forces (valid for the purposes of this paper) by Wolff et al.~\cite{Wolff2014}. Because of these overlap integrals, SBS can be symmetry-forbidden for certain optical and acoustic mode combinations (not considered here). More detailed discussions of the permitted transitions can be found in \cite{Qiu2013,WolffOptExp2014}.

Even though SBS coupling is typically considered as being dominated by bulk optical electrostriction, recent work has demonstrated that boundary optical forces play a significant role in confined geometries. These effects become extremely important in nanoscale waveguides \cite{Shin2013,Rakich2012,VanLaer2014,VanLaer2015}. The electrostrictive boundary forces arise because of the difference in optically induced stress across the waveguide-cladding boundary, while radiation pressure occurs due to momentum exchange from photons reflecting off these surfaces. Radiation pressure can also manifest within the bulk of the material \cite{Nelson1991}, but this effect is orders-of-magnitude smaller than the effects considered here. Better overlap between all these optical forces and the acoustic wave in tightly confined waveguides increases the SBS gain as well. Recent experimental work \cite{Shin2013} on linear waveguides with sub-wavelength cross-sections has demonstrated enhancement of SBS gain due to the appearance of these large boundary optical forces and the high degree of modal overlap. Based on these insights, mathematical models for computing SBS gain in linear systems have already been developed offering robust predictions for straight waveguides \cite{Qiu2013,Rakich2012,Wolff2014}, slot waveguides \cite{VanLaer2014}, and fibers \cite{Kang2010}. The action of these forces in WGRs has remained unexplored.

\begin{figure}[t]
	\makebox[\textwidth][c]{\includegraphics[width=.6\textwidth,clip=true, trim=4.2in .8in 4.5in .2in]{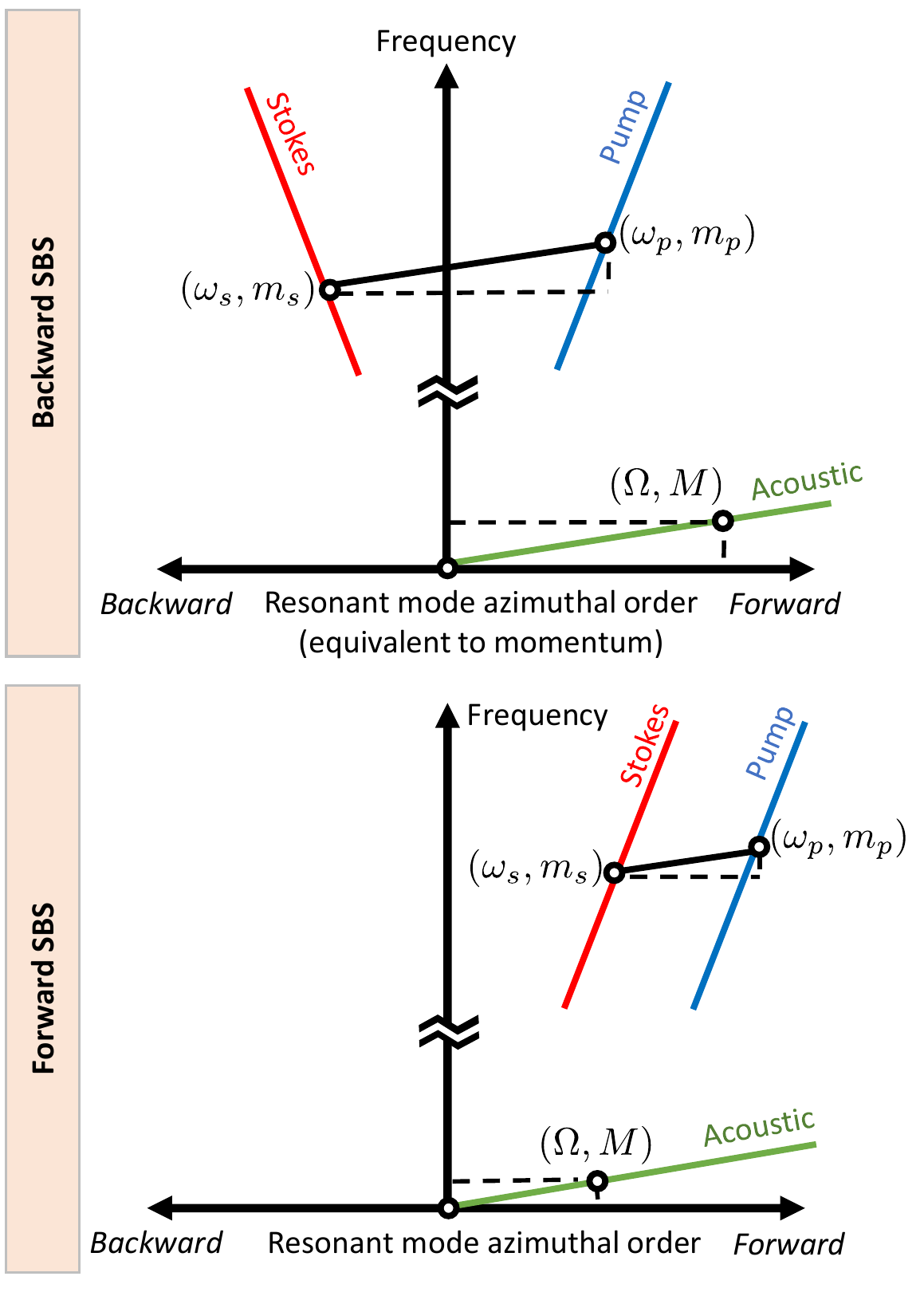}}
	\caption{\textbf{Dispersion diagrams showcasing the phase-matching requirements for SBS in a WGR.} (Top) Stokes scattering in backward SBS, and (bottom) in Forward SBS.}
	\label{fig:1}
\end{figure}

\section{The Difference Between Waveguides and\\Whispering-Gallery Resonators}
\label{sec:difference_waveguides_wgrs}

The significant field enhancement within high-finesse WGRs enables very low pump power threshold for almost any nonlinear process, from the laboratory reference frame. This is because the optical power circulating within WGRs is amplified with respect to power coupled in from the waveguide by a factor of Finesse$/2\pi$. This amplification factor can easily exceed $10^5$ for microscale silica spheres \cite{Gorodetsky1996}, shells \cite{Sumetsky2010}, and disks \cite{Lee2012}. For SBS in WGRs, both the pump and Stokes beams are resonant with comparable finesse, so the resonant enhancement of gain can be on the order of $10^{10}$ for ultra-high-Q devices. The pump finesse contributes due to resonant enhancement of the pump power, while the Stokes finesse accounts for the fact that the optical path length is finesse times longer than the resonator cavity length. Thus, from the laboratory frame of reference, WGRs also exhibit ultralow threshold SBS lasing \cite{Tomes2009,Bahl2011,Grudinin2009,Lee2012} in the {\textmu}W - mW range. This high-finesse resonant enhancement effect is not the focus of this paper, but ultimately must not be neglected.

Here we aim to show that in WGRs the SBS gain experiences additional enhancement due to surface optical forces that are ignored by traditional SBS theory. To make a fair comparison between waveguides and WGRs we will ignore the finesse factor enhancement of the pump and Stokes optical fields. Instead, we will only quantify SBS gain as a function of the optical powers for pump ($P_p$) and Stokes ($P_s$) circulating within the WGR.

In WGRs all three optical and acoustic actors participating in SBS are resonant, which enforces an integer azimuthal mode order. While the modal dispersion relationships in waveguides are usually described in $\omega-k$ space, in WGRs it is preferable to employ a $\omega-M$ space description due to this azimuthal discretization. Here, $m$ or $M$ denotes the integer azimuthal mode order i.e. the number of wavelengths around the resonator circumference, and is equivalent to the wave momentum $k$ \cite{Bahl2012Cool,Bahl2011}. The phase matching requirement for F-SBS is then expressed by the equations $\omega_p-\omega_s=\Omega$ and $m_p-m_s=M$, presented graphically in Fig.~\ref{fig:1}. In the case of B-SBS, the azimuthal relationship is modified to $m_p+m_s=M$. Here, $m_p$, $m_s$, and $M$ correspond to the pump, Stokes (scattered) and acoustic modal orders respectively.

In linear waveguides, the two optical waves can reside in a single waveguide mode whose dispersion has been suitably engineered to match the acoustic dispersion \cite{Shin2013}. In WGRs the phase matching between pump and Stokes signals is achieved at avoided crossings of two optical mode families \cite{Bahl2011}, unless anomalous dispersion occurs \cite{Matsko2009}. These optical mode shapes are well studied and easily extracted for axisymmetric geometries of WGRs \cite{Oxborrow2007,Krupka1994}. The acoustic modes can also be theoretically or computationally evaluated \cite{Matsko2009} under assumptions of circumferential periodicity using a finite element solver \cite{Bahl2012WGM}. As in linear waveguides, the evaluation of SBS gain in WGRs must incorporate the variety of optical and acoustic mode shapes, and their vector nature.

It is important to consider slight inaccuracies in phase-matching that are expected from any real system, which are contigent on the frequencies and linewidths of the three modes participating in SBS. The degree of mismatch and the resulting SBS gain spectrum can be modeled with the help of detuning parameters \cite{Kim15,Agarwal2013} and will certainly affect the achievable gain, but always scales with the line-center gain. Thus, in this paper, we consider only the maximum line-center SBS gain under the assumption that phase matching is perfectly satisfied.

While not considered here, Kerr and four-wave mixing nonlinearities \cite{Kipp2004,DelHaye2007,Moore2011,Wang2011} are known to appear at very low thresholds in high-Q resonators. These effects can become significant in silicon \cite{Shin2013} due to its stronger nonlinear response than silica, and in other highly nonlinear materials like chalcogenide glasses and ferroelectric crystals.

\begin{figure}[tb]
\makebox[\textwidth][c]{\includegraphics[width=.7\textwidth,clip=true, trim=3.1in 3.3in 4.5in 1.7in]{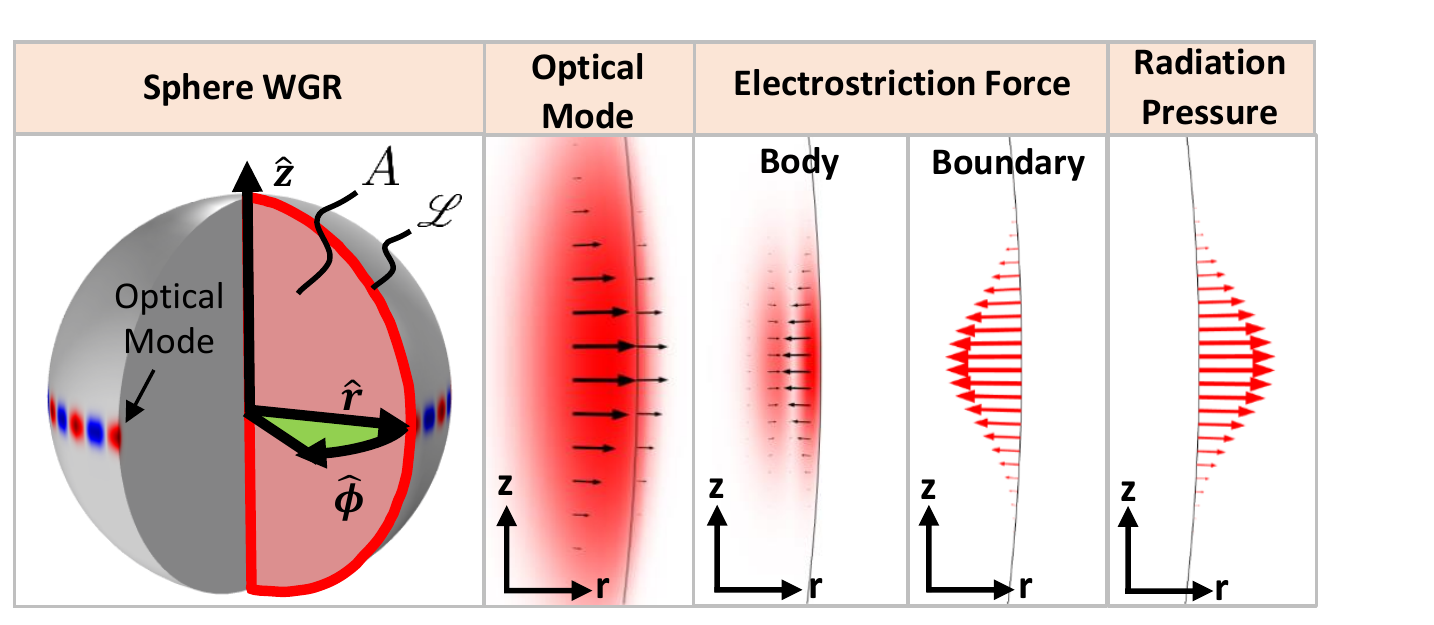}}
\caption{\textbf{Illustration of forces and whispering-gallery optical mode in a spherical WGR}. For the optical mode, color intensity depicts the magnitude of the electric field while the black arrows depict the polarization. Electrostriction body force appears in the bulk, with color indicating the force magnitude. Boundary electrostriction and radiation pressure forces appear at the free surfaces, with magnitude and direction indicated by the red arrows. }
\label{fig:2}
\end{figure}

\section{Theory of SBS Gain in WGRs}
\label{sec:wgr_theory}

Existing full vectorial models of Brillouin gain in linear waveguides \cite{Qiu2013,Rakich2012,VanLaer2014,Wolff2014,Baker2014,Kang2010} employ the Cartesian coordinate system, which is not suitable for WGR analysis. Here we follow the method of ref.~\cite{Rakich2012} to produce a vectorial formulation based on energy and particle conservation for calculating the SBS gain in axisymmetric WGRs. The optical force calculations are adapted from the theory developed in \cite{Qiu2013}. Note that geometry differences necessitate notational variations with respect to \cite{Rakich2012}. Our derivation also adds to previous work in \cite{Rakich2012} by directly defining the acoustic amplitude through the optical forces, in order to provide a formulation of SBS gain that does not require knowledge of the acoustic amplitude. This is possible under the assumption that the acoustic and optical force envelope functions have reached steady state.

We employ the following conventions: $^*$ denotes the complex conjugate, and $p$, $s$, $\Omega$ subscripts denote variables that correspond to the pump optical wave, Stokes optical wave, and acoustic wave, respectively. The cylindrical coordinate system is represented by (see Fig.~\ref{fig:2}) $\hat{r}$ denoting the radial unit vector, $\hat{z}$ denoting the axial unit vector, and $\hat{\phi}$ is the azimuthal unit vector in the counterclockwise direction. The subscripts $i$, $j$, $k$, and $l$ are used to denote directional variables.


We begin in a manner similar to \cite{Rakich2012} by defining the change in Stokes optical power $P_s$ with respect to some angular travel $\delta\phi$ (units of $rad$) as follows
\begin{equation}
\label{eq:1}
\dfrac{\delta P_s}{\delta \phi}=G_BP_pP_s-\beta_sP_s
\end{equation}
where $G_B$ (units of rad\textsuperscript{-1}W\textsuperscript{-1}) is defined as the SBS gain per angular travel averaged over an acoustic period. Here, $P_p$ is the pump optical power and the angular photon loss rate is $\beta_s$ and has units of rad\textsuperscript{-1}. Note that $P_p$ and $P_s$ are defined here as the steady-state circulating optical power within the resonator, and are thus related to the input powers via the optical finesse ($P_{res} = \frac{\omega_0 Q_{opt}}{\Delta \omega_{FSR}}P_{in}$).\par
The phonon generation rate in this system is related to the change in acoustic power per unit angle, which is equivalent to the time averaged overlap between the force and the deformation velocity over a differential volume delimited by the above $\delta\phi$. Here, the angle $\delta\phi$ covers an integer number of acoustic wavelengths, and the associated resonator volume is $\delta V$. Thus the acoustic power added to the system by the force field driving the elastic deformation wave over an integer multiple of acoustic periods can be represented by the equation
\begin{equation}
	\label{eq:2}
	\delta P_{\Omega}=\int_{\delta V}\left\langle\vec{F}_{\Omega}(\vec{r},t)\cdot\dot{\vec{u}}_{\Omega}(\vec{r},t)\right\rangle dV.
\end{equation}
Here, $\vec{F}_{\Omega}(\vec{r},t)$ is defined as the vectorial optical force density (units N/m\textsuperscript{3}) as it varies over time and space within the acoustic period-delimited volume, $\delta V$. Similarly, $\dot{\vec{u}}_{\Omega}(\vec{r},t)$ is the particle velocity field (units m/s) varying over time and space within $\delta V$. This particle velocity field is a vector field describing the velocity of individual particles in the elastic wave. Additionally, $\left\langle\cdots\right\rangle$ represents time averaging over an acoustic period, or equivalently, spatial averaging over an acoustic period.

Since particle conservation must apply, the Stokes photon flux generation rate ($\gamma_s=\frac{G_BP_pP_s}{\hbar\omega_s}$) in the resonator is equal to the phonon flux generation rate ($\gamma_{\Omega}=\frac{\delta P_{\Omega}}{\delta\phi}\frac{1}{\hbar\Omega}$). Thus we can equate $\gamma_s=\gamma_{\Omega}$ and write
\begin{equation}
	\label{eq:3}
	G_B=\dfrac{1}{\delta\phi}\dfrac{\omega_s}{\Omega}\dfrac{1}{P_pP_s}\int_{\delta V}\left\langle\vec{F}_{\Omega}(\vec{r},t)\cdot\dot{\vec{u}}_{\Omega}(\vec{r},t)\right\rangle dV.
\end{equation}
This expression is the same as that found in \cite{Rakich2012} with the exception that the differential volume is defined by the small angle $\delta\phi$ rather than a small distance $\delta z$. Following the same method as \cite{Rakich2012} we can simplify this expression to
\begin{equation}
	\label{eq:4}
	g_B=\dfrac{\omega_s}{2R\Omega}\dfrac{\bar{F}\bar{\dot{u}}}{\sqrt{P_pP_s}}\int_ARe[\mathcal{F}\cdot\mathcal{U}^*]rdrdz.
\end{equation}
The new gain variable used in Eq.~\eqref{eq:4}, $g_B$, represents the Brillouin gain over interaction distance (units m\textsuperscript{-1}W\textsuperscript{-1}) with the assumption that all interactions occur near the outermost resonator surface. The force amplitude $\bar{F}(t)$ and acoustic amplitude $\bar{\dot{u}}(t)$ have been separated from their complex unitless vectorial mode shapes ($\mathcal{F}$, $\mathcal{U}$), through the expressions $\vec{F}_{\Omega}(\vec{r},t)=\sqrt{P_pP_s}Re[\bar{F}(t)\mathcal{F}(r,z)e^{j(\Omega t-M\phi)}]$ and $\dot{\vec{u}}_\Omega(\vec{r},t)=Re[\bar{\dot{u}}(t)\mathcal{U}(r,z)e^{j(\Omega t-M\phi)}]$. The force amplitude is also normalized to remove the dependence on the optical powers. The mode shapes are integrated over the half-cross-section of the resonator, $A$ (see Fig.~\ref{fig:2}), and have been defined such that $\int_{A}\left|\mathcal{F}(r,z)\right|dA\equiv1$ and $\int_{A}\left|\mathcal{U}(r,z)\right|^2dA\equiv1$. These are both `power-normalized' integrals in the sense that they are unaffected by changes to the optical or acoustic powers. We note here that all normalizations above are performed only out to a finite radius, following the standard WGM normalizations as used in \cite{Rowland1993}, in order to avoid divergence of the integral.

We note here that a key difference arises between the gain equation for linear systems \cite{Rakich2012} and the corresponding equation in WGRs (Eq.~\eqref{eq:4}); a factor of $R$ appears in the denominator here due to the conversion from rad\textsuperscript{-1} to m\textsuperscript{-1}. This factor can be thought of as `normalizing' the radial weighting of the field within the integral.

We now remove the acoustic variable $\bar{\dot{u}}$ from Eq.~\eqref{eq:4} by considering the optical forces as a source in the acoustic wave equation. Here we model the acoustic wave as a whispering gallery mode which can be written as $\vec{u}_\Omega(\vec{r},t)=Re[\bar{u}(t)\mathcal{U}(r,z)e^{j(\Omega t-M\phi)}]$, traveling in a circumferential path along the azimuthal direction. This acoustic mode, written in terms of the particle \emph{displacement} field $\vec{u}_\Omega$, is directly related to the particle \emph{velocity} field $\dot{\vec{u}}_\Omega$ introduced previously at steady-state by $\dot{\vec{u}}=j\Omega\vec{u}$. We can thus write the acoustic wave equation as
\begin{equation}
	\label{eq:5}
	\ddot{\vec{u}}_\Omega = \nabla\cdot\frac{C}{\rho}:\nabla_s\vec{u}_\Omega+\nabla\cdot\frac{\eta}{\rho}:\nabla_s\dot{\vec{u}}_\Omega+\dfrac{\vec{F}_\Omega}{\rho}.
\end{equation}
Here $\nabla\cdot$ is the tensor divergence, $C$ is the fourth order stiffness tensor, the $:$ symbol denotes a tensor inner product, $\nabla_s$ denotes is the symmetric tensor gradient, $\rho$ is the material density, and $\eta$ is the fourth order viscosity tensor \cite{Auld1990}. In order to introduce a scalar, phenomenological damping parameter $\Gamma$ analogous to that used by Boyd~\cite{Boyd2008}, we assume the viscosity tensor is directly related to the stiffness tensor by $\eta=\frac{\Gamma}{\Omega_0}C$. At steady state the force and acoustic envelope $\bar{F}$ and $\bar{u}$ are not varying in time, allowing us to arrive at
\begin{equation}
	\label{eq:6}
	\bar{u}\mathcal{U}=\dfrac{j\bar{F}\sqrt{P_pP_s}}{\rho\Omega\Delta}\mathcal{F}.
\end{equation}
We have defined $\Delta=\Gamma+j(\Omega_0-\Omega)$ as the total loss associated with both damping and detuning from the acoustic resonance $\Omega_0$, where $\Gamma$ is the phonon loss rate in Hz. The line center gain can be calculated by setting the frequency $\Omega=\Omega_0$. We note that $\bar{u}^2$ gives the Lorentzian shape of the acoustic resonance as expected. 
Multiplying both sides by $\mathcal{U}^*$, and integrating over the cross-section, we insert Eq.~\eqref{eq:6} into Eq.~\eqref{eq:4}. We now consider only the line-center SBS gain and also introduce the mechanical quality factor, $Q_\Omega=\frac{\Omega}{\Gamma}$, to find
\begin{equation}
	\label{eq:7}
	g_B=\dfrac{Q_\Omega\omega_s}{2R\rho\Omega^2}\left(\bar{F}\int_ARe[\mathcal{F}\cdot\mathcal{U}^*]rdrdz\right)^2.
\end{equation}
It can be seen that the gain is dependent on the phonon loss rate in the same manner as a linear waveguides \cite{Boyd2008,Rakich2012}. The nonlinear dependence on the force can thus be seen to arise from the relation of the acoustic amplitude to the optical force ($\bar{\dot{u}}\propto\bar{F}\sqrt{P_pP_s}\int_ARe[\mathcal{F}\cdot\mathcal{U}^*]dA$).

Finally, the optical force term within Eq.~\eqref{eq:7} is comprised of electrostriction $\bar{F}_{ES}$ (which has both body and boundary components) and radiation pressure $\bar{F}_{RP}$, such that $\bar{F}=\bar{F}_{ES,body}+\bar{F}_{ES,boundary}+\bar{F}_{RP}$. The line-center SBS gain $g_B$ in WGRs is thus given by
\begin{multline}
	\label{eq:8}
	g_B=\dfrac{Q_\Omega\omega_s}{2R\rho\Omega^2}\left(\bar{F}_{ES,body}\int_ARe[\mathcal{F}_{ES,body}\cdot\mathcal{U}^*]rdrdz+\right.\\
	\bar{F}_{ES,boundary}\int_\mathcal{L}Re[\mathcal{F}_{ES,boundary}\cdot\mathcal{U}^*]rdl+\\
	\left.\bar{F}_{RP}\int_\mathcal{L}Re[\mathcal{F}_{RP}\cdot\mathcal{U}^*]rdl\right)^2.
\end{multline}
As can be seen, the gain depends on the integral of the phonon generation rate over the half-cross-section $A$ of the resonator for the body force, and over the line $\mathcal{L}$ bounding the half-cross-section for the boundary forces (see Fig.~\ref{fig:2}). Again, we note that this gain equation differs from its linear counterpart in \cite{Rakich2012} since the acoustic power is weighted by the radial location $R$ of the wave.

\subsection*{Evaluating the optical forces}
\label{subsec:optical_forces}

We now require analytical expressions for the individual surface and boundary forces in terms of the optical fields. Here we can assume that the two optical signals are confined to a narrow region within a few microns of the resonator surface. The azimuthal component of each vector thus points in the direction of wave propagation, and the radial and axial components are transverse to propagation, similar to a plane wave.\par
The electrostriction stress on the material is related to the optical field through the photoelastic tensor as shown in \cite{Qiu2013}. Since silica is an amorphous isotropic material, the electrostriction stresses in our cylindrical coordinate system can be written through the expression

\begin{equation}
\label{eq:9}
    \begin{bmatrix}
    	\sigma_{rr}\\
    	\sigma_{zz}\\
    	\sigma_{\phi\phi}\\
    	\sigma_{rz}\\
    	\sigma_{r\phi}\\
    	\sigma_{z\phi}
    \end{bmatrix}
	= -\dfrac{1}{4}  \epsilon_0  n^4 \bar{P}
    \begin{bmatrix}
    	2E_{p,r}E^*_{s,r}\\
    	2E_{p,z}E^*_{s,z}\\
    	2E_{p,\phi}E^*_{s,\phi}\\
    	E_{p,r}E^*_{s,z}+E_{p,z}E^*_{s,r}\\
    	E_{p,r}E^*_{s,\phi}+E_{p,\phi}E^*_{s,r}\\
    	E_{p,z}E^*_{s,\phi}+E_{p,\phi}E^*_{s,z}
    \end{bmatrix}
\end{equation}
where the photoelastic tensor $\bar{P}$ is
\begin{equation}
	\bar{P} =     
	\begin{bmatrix}
    		p_{11} & p_{12} & p_{12} & 0 & 0 & 0\\
    		p_{12} & p_{11} & p_{12} & 0 & 0 & 0\\
    		p_{12} & p_{12} & p_{11} & 0 & 0 & 0\\
    		0 & 0 & 0 & p_{44} & 0 & 0\\
    		0 & 0 & 0 & 0 & p_{44} & 0\\
    		0 & 0 & 0 & 0 & 0 & p_{44}\\
	\end{bmatrix}
\end{equation}
Here, $\sigma_{ij}$ denotes the components of the electrostriction stress, $n$ is the effective refractive index of the resonator material, the $p_{ij}$ terms are the components of the photoelastic tensor, the $p$ and $s$ subscripts denote the pump and Stokes fields, and the $r$, $z$, and $\phi$ subscripts denote the directional components. Note that $\sigma_{ij}=\sigma_{ji}$ for the purposes of this paper.\par
The electrostriction body force is then obtained from the summed gradients of the normal ($i=j$) and transverse ($i\neq j$) stresses as shown in \cite{Qiu2013}, which in our cylindrical coordinate system are expressed as
\begin{subequations}
\label{eq:10}
\begin{equation}
\vec{F}_r^{ES,body}=-\dfrac{\partial}{\partial r}\sigma_{rr}-\dfrac{\partial}{\partial z}\sigma_{rz}-j\dfrac{M}{r}\sigma_{r\phi}
\end{equation}
\begin{equation}
\vec{F}_z^{ES,body}=-\dfrac{\partial}{\partial z}\sigma_{zz}-\dfrac{\partial}{\partial r}\sigma_{zr}-j\dfrac{M}{r}\sigma_{z\phi}
\end{equation}
\begin{equation}
\vec{F}_\phi^{ES,body}=-j\dfrac{M}{r}\sigma_{\phi\phi}-\dfrac{\partial}{\partial r}\sigma_{\phi r}-\dfrac{\partial}{\partial z}\sigma_{\phi z}.
\end{equation}
\end{subequations}
An additional electrostriction boundary force exists at material interfaces, arising from the difference in electrostriction stresses across the media. Here we assume that the resonator is immersed in air, which supports negligible electrostriction stress in comparison to the resonator material (it is zero in a vacuum environment). The electrostriction boundary force can then be written as
\begin{subequations}
	\label{eq:11}
	\begin{equation}
	\vec{F}_r^{ES,boundary}=\sigma_{rr}\hat{n}_r+\sigma_{rz}\hat{n}_z
	\end{equation}
	\begin{equation}
	\vec{F}_z^{ES,boundary}=\sigma_{zz}\hat{n}_z+\sigma_{zr}\hat{n}_r
	\end{equation}
	\begin{equation}
	\vec{F}_\phi^{ES,boundary}=\sigma_{\phi r}\hat{n}_r+\sigma_{\phi z}\hat{n}_z.
	\end{equation}
\end{subequations}
Here, $\hat{n}_{r,z}$ are the radial and axial components of the boundary surface normal vector $\hat{n}$, and are transverse to wave propagation.\par
The final contributing force is radiation pressure, which arises from the Maxwell Stress Tensor \cite{Panofsky2005}. When considering dielectric systems without free charges, as is the case for this study, radiation pressure is localized where the dielectric constant gradient is nonzero. When the system consists only of domains of homogeneous media, radiation pressure exists only on material boundaries. Again considering a resonator immersed in air or vacuum, the boundary radiation pressure force can be written as
\begin{equation}
\label{eq:12}
\vec{F}^{RP}=\dfrac{\epsilon_0\epsilon}{2}(\epsilon-1)E_{p,n}E^*_{s,n}\hat{n}
\end{equation}
where $\epsilon$ denotes the relative permittivity of the resonator, the $n$ subscript denotes the component of the electric field which is normal to the boundary, and $\hat{n}$ denotes the boundary normal vector pointing outwards from the resonator. Note that electrostriction and radiation pressure, due to their distinct physical origins \cite{Juretschke1997,Lee2005}, must be included separately as we have done above.

At first sight the two boundary forces normal to the resonator surface may seem mathematically orthogonal to the azimuthal propagation of the acoustic wave and unable to contribute to the generation of azimuthal phonons. However, we remind the reader that these optical forces are a consequence of the spatiotemporal interference between the pump and Stokes optical fields. These forces oscillate at the pump-Stokes beat frequency, and also \emph{spatially propagate} with the momentum-vector difference between them. Due to the defined SBS phase-matching relationship, this spatiotemporally varying force precisely matches the frequency and momentum vector of the acoustic wave, thus co-propagating with the acoustic wave. Since all surface acoustic waves involve some normal displacement of the material surface, boundary-normal forces do actuate these waves. In other words, boundary forces have non-zero overlap integral with the traveling wave acoustic mode and are able to drive the coherent generation of phonons in the system. This interpretation is well known from prior work on SBS enhancement in nanoscale waveguides and also in resonators \cite{Bahl2012Cool}.

\subsection*{Scalar Brillouin Gain}
\label{subsec:scalar_gain}
In order to provide a comparison benchmark for the vectorial SBS gain, we now evaluate the scalar SBS gain. Here we assume that a longitudinal acoustic wave is driven only by bulk electrostriction as considered in standard SBS theory \cite{Boyd2008}. Additionally, the relation between the optical forces and acoustic wave is further simplified by only considering the contribution of the transverse optical fields to the electrostriction stress; this simplification reduces the photoelastic tensor to a scalar ($p_{12}$). After making these simplifications, the scalar gain can be written as
\begin{equation}
\label{eq:13}
g_B=\dfrac{M}{m_s}\dfrac{n^7p_{12}^2\omega_s^2}{2c^3\rho\nu}\dfrac{Q_\Omega}{\Omega}\dfrac{\eta}{R}
\end{equation}
where $c$ is the speed of light and we have introduced $\eta$ as the spatial overlap integral. The spatial overlap integral describes the efficiency with which the optical forces drive the acoustic wave and can be evaluated here as $\eta=\left(\int_ARe[\mathcal{F}\cdot\mathcal{U}^*]rdrdz\right)^2$.\par
For smaller resonators modal confinement is improved, thus the relative overlap between optical and acoustic fields tends to increase. As a result, small resonators will naturally exhibit a larger SBS gain. However, this trend cannot always be predicted because of the unique shapes of optical and acoustic modes. For instance, the symmetric Lamb wave modes considered in the next section have sufficient acoustic dispersion variation with decreasing resonator thickness to significantly decrease the scalar gain because of the increase in acoustic frequency (gain is quadratically dependent on acoustic frequency $\Omega$ as shown in Eq.~\eqref{eq:8}).

\section{Brillouin Gain Calculations and Discussion}
\label{sec:discussion}

Using the formalism established in Section~\ref{sec:wgr_theory}, we now calculate the full vectorial Brillouin gain (Eqs.~\eqref{eq:8}-\eqref{eq:12}) and the scalar gain (Eq.~\eqref{eq:13}) for some cases of interest. As we shall see, significant deviations from scalar theory are predicted by the vectorial numerical calculations that incorporate the boundary forces. In particular, giant SBS gain enhancement up to $10^4$ greater than scalar theory, and significant suppression of SBS gain by as much as $10^4$ times, are predicted in specific cases.

\subsection*{Cases Considered and Key Assumptions}

\begin{figure*}
	\centering
	\includegraphics[width=0.85\textwidth,clip=true,trim=4.3in 0in 1in .3in]{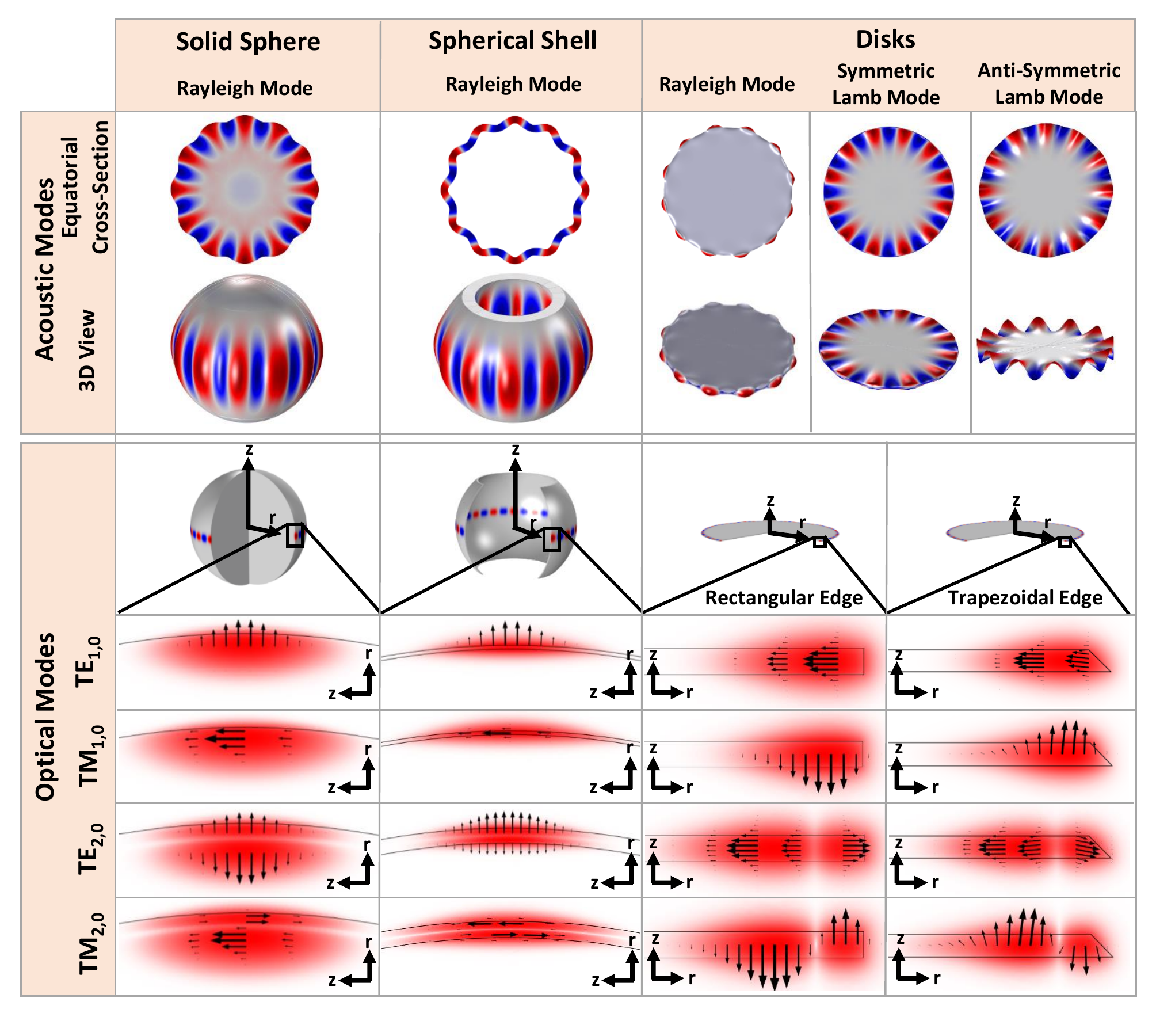}
	\caption{\textbf{Examples of acoustic and optical modes used in this work}. For acoustic modes, we present equatorial cross-sections and full 3D views of the acoustic modes with exaggerated deformation. Red and blue denote the magnitude and direction of the displacement. For optical modes, we present azimuthal cross-sections of the optical modes in conjunction for each WGR. Four lowest optical mode families are examined: TE$_{1,0}$, TM$_{1,0}$, TE$_{2,0}$, and TM$_{2,0}$. The color intensity represents the magnitude while the arrows represent the polarization of the electric field. TE$_{1,0}$ is radially polarized and concentrated in a single antinode. TM$_{1,0}$ is axially polarized and concentrated in a single antinode. TE$_{2,0}$ is radially polarized and concentrated in two antinodes, distributed radially. TM$_{2,0}$ is axially polarized and concentrated in two antinodes, distributed radially.}
\label{fig:3}
\end{figure*}

Our search here is focused on F-SBS only. This is because B-SBS involves extremely high-frequency, short-wavelength acoustic modes that are spectrally dense \cite{Tomes2009,Bahl2011Characterization}, preventing accurate identification in experiments on large WGRs. However, we stress that the theoretical foundations of Section~\ref{sec:wgr_theory} remain valid even for B-SBS. The validity of Eq.~\eqref{eq:8} for both F-SBS and B-SBS leads to the expectation that the trends predicted here (such as scaling of gain with resonator geometry) will also appear in the B-SBS case. Overall lower SBS gain should be expected for B-SBS due to the inverse-quadratic dependence of the gain on acoustic frequency $\Omega$ and linear dependence on acoustic $Q_\Omega$, slightly counter-balanced by the decreased acoustic mode volume which improves modal overlap. Additionally, since F-SBS always has higher gain (phonon lifetimes are significantly longer at lower frequencies), we are very interested in how much this gain can be enhanced.

Four basic WGR geometries (Fig.~\ref{fig:3}) are considered in this study: solid sphere, spherical shell, rectangular disk, and trapezoidal disk. Solid microspheres are easily made on the tips of silica fibers using reflow or arc discharge processes \cite{Braginsky1989}. Spherical shells encompass the microbottle \cite{Sumetsky2004} and microbubble \cite{Sumetsky2010} WGR categories. Trapezoidal disks can be fabricated through wet-etching \cite{Lee2012}, while rectangular cross-section disk resonators are fabricated through directional ion-etching \cite{Borselli2005}.

While many high-order optical WGMs exist in such resonators, here we constrain the analysis by considering only four lowest-order TE-like (radially polarized) and TM-like (axially polarized) modes \cite{Gorodetsky1999,Little1999}, all of which are illustrated in Fig.~\ref{fig:3}. The modal subscript numbering convention used here is common to WGRs \cite{Little1999} where the first numeral denotes the number of anti-nodes in the radial direction and the second numeral denotes the number of nodes in the axial direction. We selected the azimuthal mode order $m_p$, $m_s$ to place the optical frequencies around the $1550$ nm telecom band ($\sim193$ THz) for each case to facilitate comparison. Pairwise combinations of these simple modes are employed as the two (pump and Stokes) optical modes in the computational analysis of the SBS gain.

\begin{figure*}
	\centering
	\includegraphics[width=1\textwidth,clip=true,trim=2.8in 1.8in 3.7in 0in]{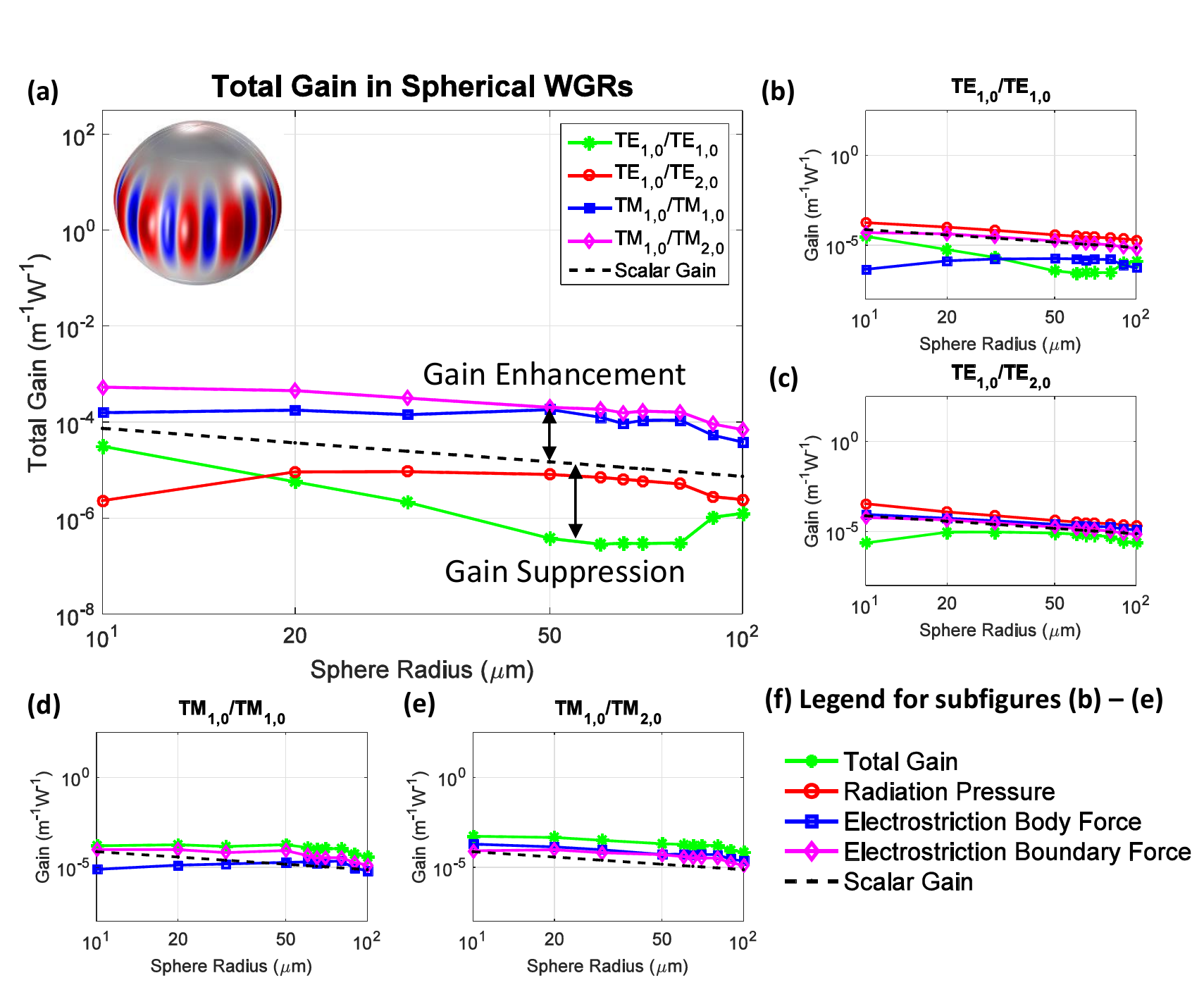}
	\vspace{-0.2in}
	\caption{\textbf{SBS gain computed in sphere WGRs with Rayleigh acoustic waves ($M=12$) exhibits both enhancement and suppression beyond scalar theory}. (a) The total gain including all forces. (b)-(e) Gain contributions of individual optical forces if they were considered independently. The SBS gain expected from scalar theory is shown for comparison on each plot. Note that radiation pressure does not appear in (d)-(e) since it is many orders of magnitude smaller than the other optical forces. (f) Legend for subfigures (b)-(e).}
	\label{fig:4}
\end{figure*}

The third actor in microresonator SBS is the acoustic WGM. Multiple surface acoustic wave mode families are considered in our analysis (Fig.~\ref{fig:3}), as they are known to exist in WGRs \cite{Bahl2012WGM} and have been experimentally observed as well \cite{Bahl2012Cool,Bahl2013,Bahl2011}. Rayleigh waves (radial displacement) are considered in all the WGRs that we consider in this analysis, symmetric Lamb waves (extensional) are considered in both disk WGR configurations, and anti-symmetric Lamb waves (flexural) are considered in the trapezoidal disk WGR configuration. In each case, the acoustic mode that satisfies SBS phase-matching between the selected optical modes must follow $M=m_p-m_s$ as previously described in Section~\ref{sec:difference_waveguides_wgrs}. However, since each resonator's size and dispersion determines $m_p$ near $1550$ nm, we risk having no standard point of comparison between the resonator geometries. We thus constrain our search to only the lowest transverse order acoustic modes, while the azimuthal mode order is fixed at $M=12$ as a standard comparison point commensurate with past experimental observations \cite{Bahl2012Cool,Bahl2013}. The pump azimuthal mode order $m_p$ is then selected as described above, and the availability of a Stokes optical mode with correct $m_s$ is assumed. This value of $M$ was chosen since numbers in this range have been measured experimentally \cite{Bahl2012Cool,Bahl2013,Bahl2011}. 

In the cases where we consider intra-modal scattering (example - the TE$_{1,0}$/TE$_{1,0}$ case) the scattering is taking place between the $N$\textsuperscript{th} and ($N+M$)\textsuperscript{th} azimuthal order optical modes having the TE$_{1,0}$ mode shape. Here, $N$ is a large integer depending on the circumference of the resonator, while $M=12$ is fixed. In each calculation, we compute the nearest integer $N$ based on the resonator size. The same applies to the inter-modal scattering pairs (example - the TE$_{1,0}$/TE$_{2,0}$ case) where the scattering takes place between the $N$\textsuperscript{th} TE$_{1,0}$ mode and the ($N+M$)\textsuperscript{th} TE$_{2,0}$ mode. We also note that the optical forces generated in the reversed case of ($N+M$)\textsuperscript{th} TE$_{1,0}$ mode and the $N$\textsuperscript{th} TE$_{2,0}$ mode are nearly the same as long as $M\ll N$, since in such a case the mode shapes do not change appreciably. This symmetry can be seen in Eqs.~\eqref{eq:9}~and~~\eqref{eq:12} in the evaluation of optical stresses and forces.

In order to set simulation parameters, and to draw comparisons to a commonly used optics platform, we have chosen to examine silica WGRs. We note, however, that the overall scaling trends observed here are not specific to any material platform as they are primarily a geometrical effect. For the purpose of our numerical computations, the mechanical quality factor is assumed to be $10^4$, selected based on experimental observations \cite{Bahl2013,Bahl2011}. In smaller resonators, the acoustic frequency for an $M=12$ mode increases, resulting in a shorter phonon lifetime for the same Q-factor. The photoelastic constants for silica \cite{Donadio2003} are set at $p_{11} = 0.125$, $p_{12} = 0.278$, and $p_{44} = -0.073$ for evaluating Eq.~\eqref{eq:9}.

Mode polarization and resonator geometric scale are seen below to be the primary drivers for the observed gain enhancement and suppression. For instance, smaller resonators tend to have increased boundary forces and better overlap between the acoustic mode and optical forces, resulting in higher gain. Brillouin gain is suppressed in cases where radiation pressure force approximately balances electrostriction. We also observe that the gain predicted by scalar theory underestimates the gain predicted by vectorial theory, especially for smaller resonator scales where the largest gain enhancements are observed.


\subsection*{Brillouin Gain in Silica Microspheres}

We first examine the SBS gain in spherical silica WGRs (see Fig.~\ref{fig:3}) since they are extensively used in microresonator studies and are easy to fabricate. The simple spherical geometry also helps bring out the relation of the opto-acoustic coupling to the resonator scale. Brillouin lasing has been experimentally demonstrated in microspheres before \cite{Bahl2012Cool,Bahl2011,Tomes2009,Kim15}, although the relative contributions of radiation pressure and electrostriction have never been analyzed previously in the resonator case. In Fig.~\ref{fig:4} we present the computed SBS gain in the Rayleigh acoustic wave case for sphere radii ranging from $10\,-\,100$ {\textmu}m using both scalar and vectorial calculations.

\begin{figure*}
	\centering
	\includegraphics[width=1\textwidth,clip=true,trim=3in 1.3in 3in .7in]{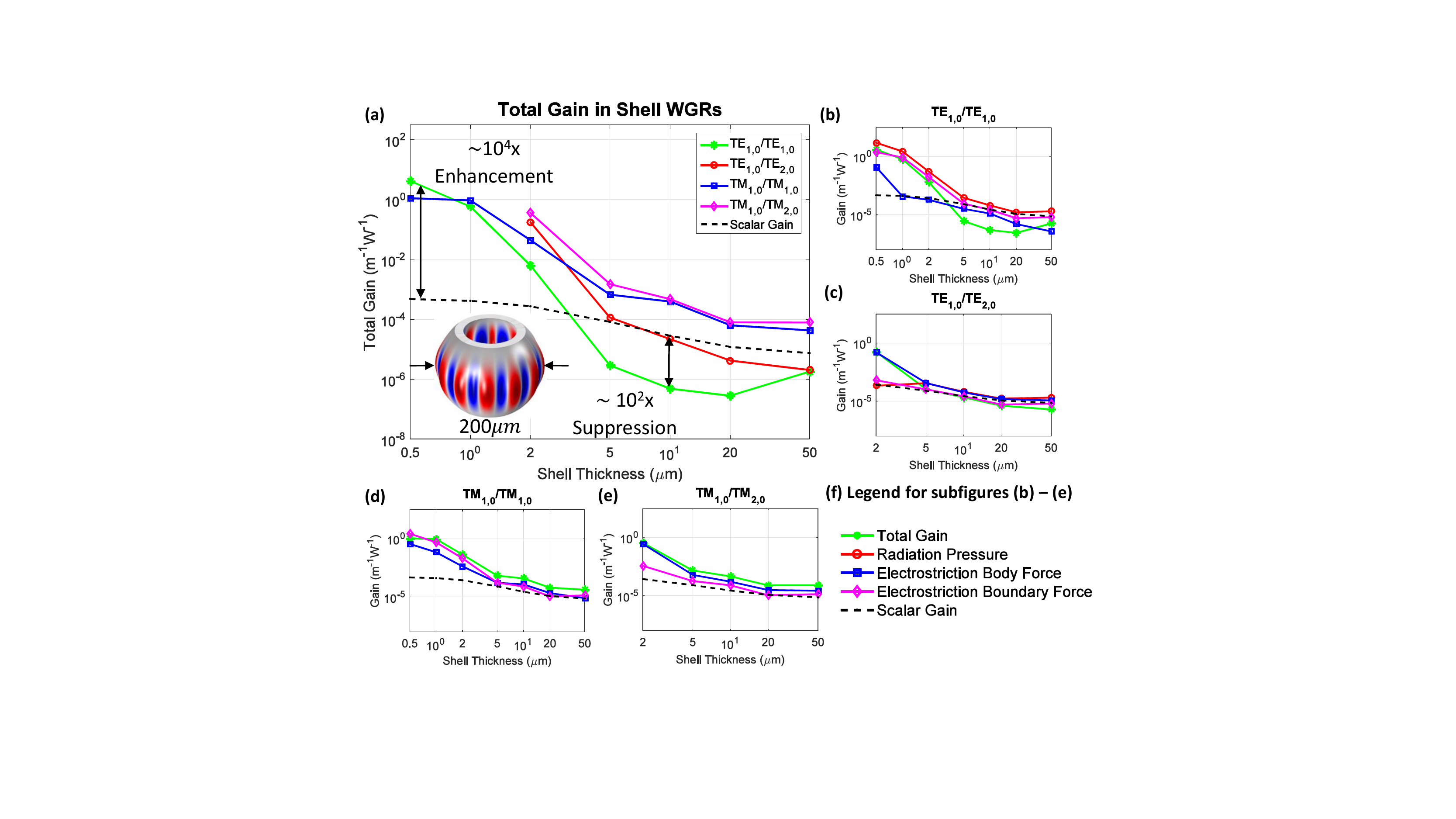}
	\caption{\textbf{SBS in shell-type WGRs (bubbles, bottles) exhibits very significant gain enhancement for the Rayleigh acoustic wave case ($M=12$)}. The causes for the gain enhancement are discussed in Fig.~\ref{fig:6}. (a) The total gain including all forces. (b)-(e) Gain contributions of individual optical forces if they were considered independently. (f) Legend for subfigures (b)-(e).}
	\label{fig:5}
\end{figure*}

We immediately note that while the scalar electrostriction-only theory approximately predicts the trend shown in Fig.~\ref{fig:4}, it over-estimates or under-estimates the full vectorial calculation by an order of magnitude in some cases. However, the scalar gain does roughly predict the contributions from each each force individually when the optical modes are relatively well confined inside the resonator (large sphere diameters), as seen in Fig.~\ref{fig:5}(b)-(e). Lowered optical confinement (for smaller sphere diameters) causes the appearance of large boundary forces and affects the overlap integral in Eq.~\eqref{eq:13}. These additional forces significantly deviate the gain from the scalar estimate since the vectorial gain is evaluated through a nonlinear combination of all contributing forces (see Eq.~\eqref{eq:8}).

As mentioned previously, radiation pressure and electrostriction can oppose each other (see Fig.~\ref{fig:2}) in certain cases. Brillouin gain suppression is seen to occur for TE modes (Fig.~\ref{fig:4}(b)-(c)) when radiation pressure and electrostriction are of approximately equivalent magnitude. However, no gain suppression occurs for TM modes (Fig.~\ref{fig:4}(d)-(e)) because the electric field component normal to the surface is very small, generating negligible radiation pressure contribution as per Eq.~\eqref{eq:12}.

These results immediately support our main assertion that SBS gain enhancement does not require nanoscale or sub-wavelength features as previously proposed \cite{Qiu2013,Shin2013,Rakich2010,Rakich2012}. SBS gain enhancement can occur whenever optical fields are confined along a surface and have good overlap with the acoustic mode. Furthermore, SBS is not solely driven by bulk electrostriction in resonators as has been thought until now, but is instead significantly affected by boundary optical forces as well.

\subsection*{Brillouin Gain in Silica Microshells}

We now consider gain in spherical shell WGRs of the form shown in Fig.~\ref{fig:3}. Computed gains for the Rayleigh acoustic wave case are presented in Fig.~\ref{fig:5} where the shell thickness is varied between $0.2\,-\,10$ {\textmu}m using a fixed radius of $100$ {\textmu}m. We immediately note giant gain enhancement ($\sim10^4\times$) for thin shells, which becomes significant when the shell thickness is on the order of a few optical wavelengths ($<5$ {\textmu}m). Since all the modes are confined near the outer WGR surface, thicker shells tend to approximate simple spheres, and the computed gain also converges to the result for a sphere (compare Fig.~\ref{fig:4}(a) and Fig.~\ref{fig:5}(a)). A notable feature is the significant gain suppression ($\sim10^2\times$) in the TE$_{1,0}$/TE$_{2,0}$ mode pair for the specific case of a $10$ {\textmu}m thick shell. Overall, it can be seen that the scalar computation is a poor predictor of Brillouin gain for thin shell WGRs.\par

\begin{figure*}
	\makebox[\textwidth][c]{\includegraphics[width=1.2\textwidth, clip=true,trim=2.3in 1.6in 2in 2in]{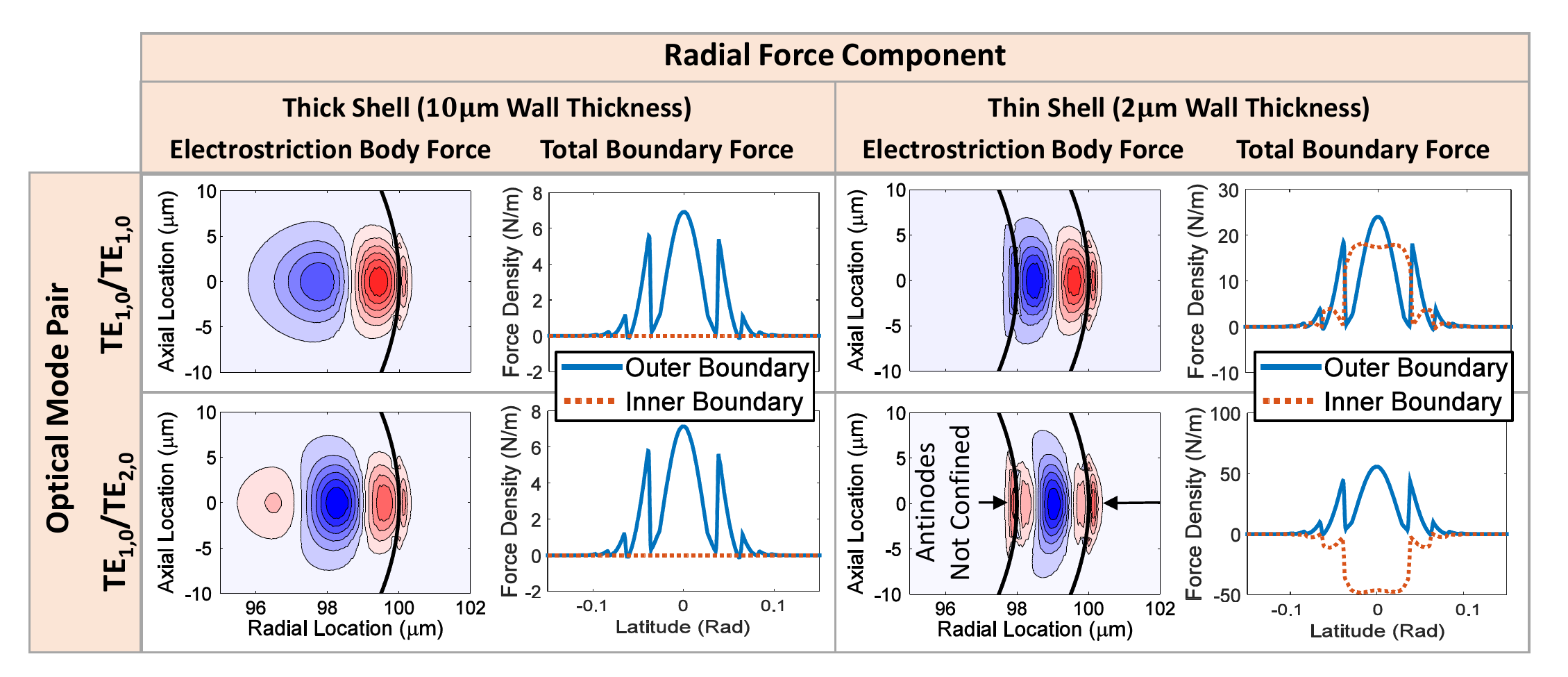}}
	\caption{\textbf{Gain enhancement in shells with Rayleigh acoustic WGMs arises due to boundary and body forces when shell thickness approaches the optical wavelength}. For the body force, red indicates force directed radially outwards and blue indicates force directed radially inwards. An intra-modal pair and an inter-modal pair are chosen in order to highlight the different force distributions in thin ($2$ {\textmu}m) and thick ($10$ {\textmu}m) shells. The force distribution shown external to the resonator does not exist but is shown only to indicate the presence of optical fields.}
	\label{fig:6}
\end{figure*}

For shell WGRs we find that there are two distinct mechanisms of gain enhancement. The intra-modal combinations (TE$_{1,0}$/TE$_{1,0}$ and TM$_{1,0}$/TM$_{1,0}$) experience gain enhancement due to large optical boundary forces, while the inter-modal combinations (TE$_{1,0}$/TE$_{2,0}$ and TM$_{1,0}$/TM$_{2,0}$) experience large gains due to the body force. These enhancement mechanisms are evidenced in Fig.~\ref{fig:6}.

\textbf{Enhancement from boundary forces for \textbf{intra}-modal scattering}: 
Let us consider the radial component of the combined boundary force for the TE$_{1,0}$/TE$_{1,0}$ mode pair (Fig.~\ref{fig:6}, top row). When the shell wall thickness decreases, the optical modes begin to interact with the inner boundary, and the boundary forces thus generated start contributing to the Brillouin gain. Typically, higher interfacial fields create stronger boundary forces and increased Brillouin gain is expected as in the case of nanoscale waveguides. In the mode combination studied here, the forces generated on the inner boundary oppose the forces on the outer boundary, resulting in a lower contribution to the gain.

\textbf{Enhancement from body force for \textbf{inter}-modal scattering}: 
Let us now consider the radial component of the electrostriction body force for the TE$_{1,0}$/TE$_{2,0}$ mode pair (Fig.~\ref{fig:6}, bottom row). For this case three regions of body force concentration are observed, which alternate between force directed radially outwards and radially inwards. These opposing forces approximately cancel, and this self-cancellation results in a reduced gain contribution from the total body force. For thinner resonators, however, the resonator does not confine the optical fields sufficiently. The central body force concentration then is the primary contributor to electrostriction within the resonator material and the above self-cancellation effect is reduced. The gain in thin shells is therefore enhanced beyond the value for thicker shells and spheres. The intra-modal combinations, on the other hand, exhibit only two opposing force concentrations such that the self-cancellation effect persists.

The significant ($\sim10^2\times$) suppression of gain in the TE$_{1,0}$/TE$_{2,0}$ mode pair for $10$ {\textmu}m thick shells is caused by the comparable and opposite radiation pressure and electrostriction forces in the same manner as described previously in spheres. It can be speculated that an exact choice of geometrical parameters could also lead to perfect cancellation and thus giant suppression of the SBS gain beyond the demonstrated value.

\begin{figure*}
	\centering
	\includegraphics[width=1\textwidth,clip=true,trim=3in 1.7in 3.2in .5in]{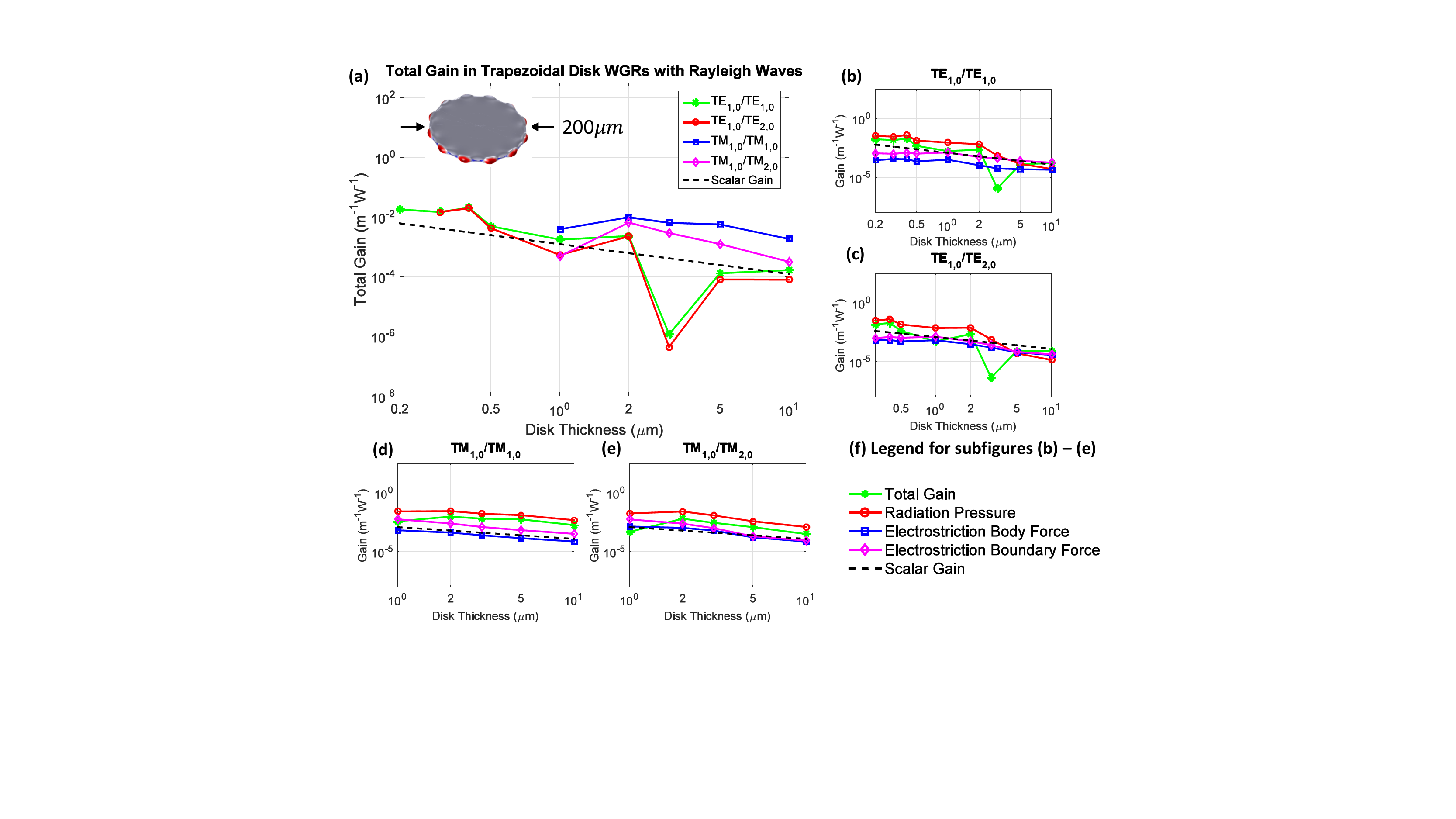}
	\caption{\textbf{SBS in trapezoidal disk WGRs with Rayleigh waves ($M=12$) is well approximated by the scalar gain}. TE mode pairs exhibit a sudden rise in total gain in disks of thickness $2\,-\,5$ {\textmu}m. (a) The total gain including all forces. (b)-(e) Gain contributions of individual optical forces if they were considered independently. (f) Legend for subfigures (b)-(e). See also Fig.~\ref{fig:8}. \vspace{0.5cm}}
	\label{fig:7}
\end{figure*}

\subsection*{Brillouin Gain in Silica Trapezoidal Microdisks}

Microdisk resonators with trapezoidal cross-section (Fig.~\ref{fig:3}) are also widely employed in photonic microsystems \cite{Lee2012,Papp2014}. For SBS gain calculations, we consider three acoustic WGM families: Rayleigh waves, symmetric Lamb waves, and anti-symmetric Lamb waves (see Fig.~\ref{fig:3}). We consider disk thickness in the range of $0.2\,-\,10$ {\textmu}m using a fixed radius of $100$ {\textmu}m. Anti-symmetric Lamb waves are seen to have the largest absolute gain ($\sim10^2$ m\textsuperscript{-1}W\textsuperscript{-1}) while the largest gain enhancement ($>10^2\times$ beyond scalar theory) is predicted for symmetric Lamb waves.

\textbf{Rayleigh acoustic WGMs}: These modes are characterized by a large radial deflection at the outer edge surface of the resonator. The results of scalar and full numerical computation in this case are presented in Fig.~\ref{fig:7}. Unlike solid spheres and spherical shells, we predict that the gain of mode pairs of both TE and TM polarizations is primarily driven by radiation pressure. The axially polarized (TM) modes are able to drive a radially directed acoustic wave by means of radiation pressure because the angled edge of the trapezoidal disk breaks symmetry.\par

\begin{figure}
	\centering
	\includegraphics[width=0.6\textwidth,clip=true,trim=4in 2.6in 4.5in 1in]{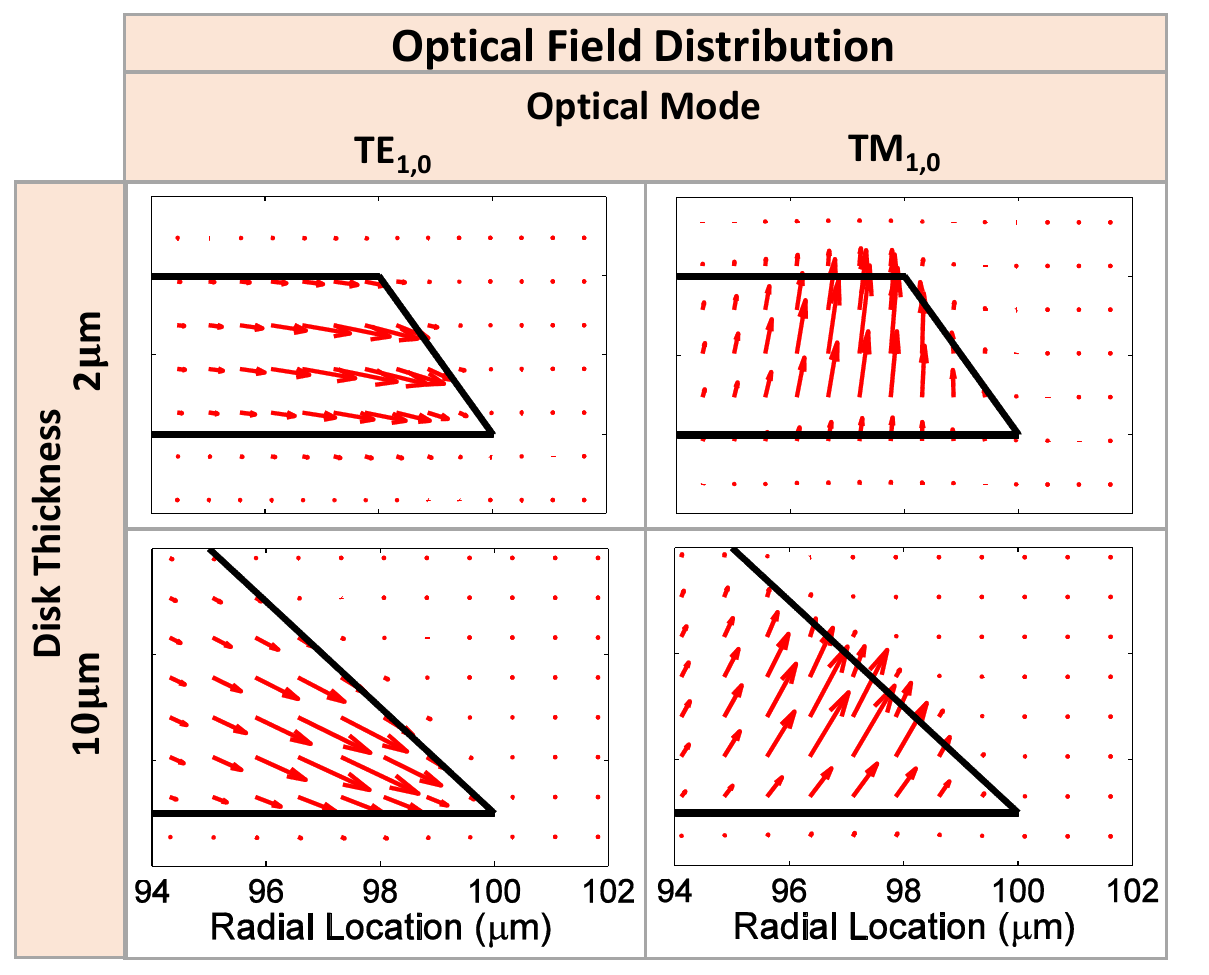}
	\caption{\textbf{The sudden rise of gain in trapezoidal disk WGRs is attributable to the location and polarization of the optical modes}. The lowest order mode of each polarization is shown in thin ($2$ {\textmu}m) and thick ($10$ {\textmu}m) disks. The red arrows indicate the direction and magnitude of the in-plane electric field.}
	\label{fig:8}
\end{figure}

For thick disks TM polarized optical mode pairs show much larger radiation pressure contribution than TE polarized mode pairs. This is because the TE mode boundary-normal E-fields at the outer edge are very small (Eq.~\eqref{eq:12}), as is simulated in Fig.~\ref{fig:8} (bottom left). In contrast, the TM modes exhibit larger boundary normal fields (Fig.~\ref{fig:8} bottom right).
For thinner disks ($<2$ {\textmu}m) that are comparable to the optical wavelength, the TE optical modes have a larger boundary-normal E-field component at the outer edge, as is simulated in Fig.~\ref{fig:8} (top left). This increases the radiation pressure contribution for the smaller size scales, with the Brillouin gain transition for Rayleigh modes being easily visible in Fig.~\ref{fig:7}(b)(c).\par

\begin{figure*}
	\centering
	\includegraphics[width=1\textwidth,clip=true,trim=3.1in 1.3in 3.2in 0.85in]{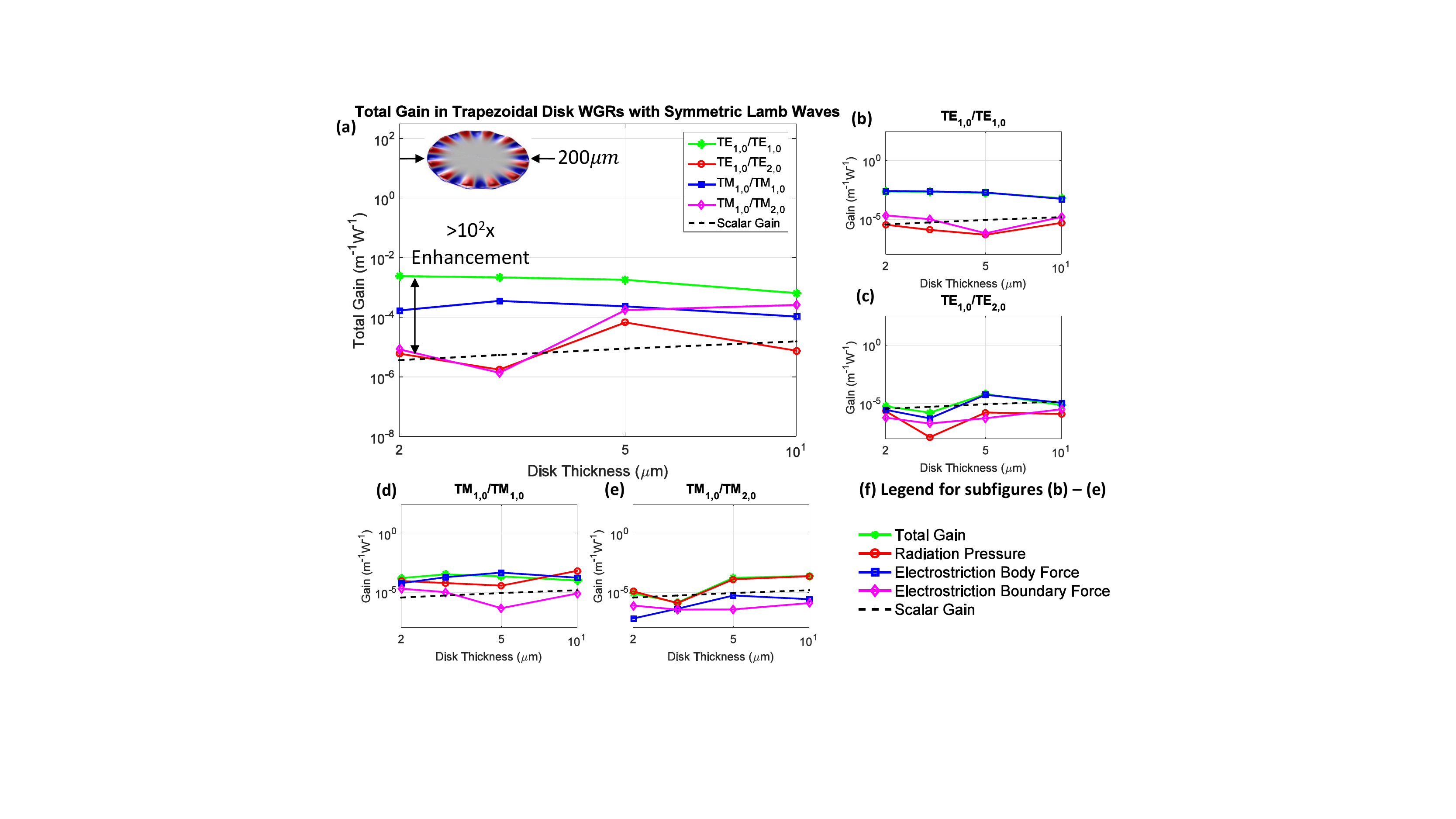}
	\vspace{-0.15in}
	\caption{\textbf{SBS in trapezoidal disk WGRs with symmetric Lamb waves ($M=12$) has low total gain but shows enhancement over scalar theory}. Disks of thickness smaller than $2$ {\textmu}m were not tested since symmetric Lamb waves transition to longitudinal waves in thin plates. Acoustic dispersion of these symmetric Lamb modes is responsible the downward trend of scalar gain for thinner disks. (a) The total gain including all forces. (b)-(e) Gain contributions of individual optical forces if they were considered independently. (f) Legend for subfigures (b)-(e).}
\label{fig:9}
\end{figure*}

\textbf{Symmetric Lamb wave WGMs}: These modes are characterized by mirrored axial deflection along the thickness axis of the disk \cite{Auld1990} as illustrated in Fig.~\ref{fig:3}. The scalar and full numerical computation results for this case are shown in Fig.~\ref{fig:9}. For very thin disks the distinction between symmetric Lamb waves and longitudinal waves is no longer apparent \cite{Pilarski1993}, so we only calculated gains for thickness ranging over $2\,-\,10$ {\textmu}m.\par

\begin{figure*}
	\centering
	\includegraphics[width=1\textwidth,clip=true,trim=3in 1.3in 3in .85in]{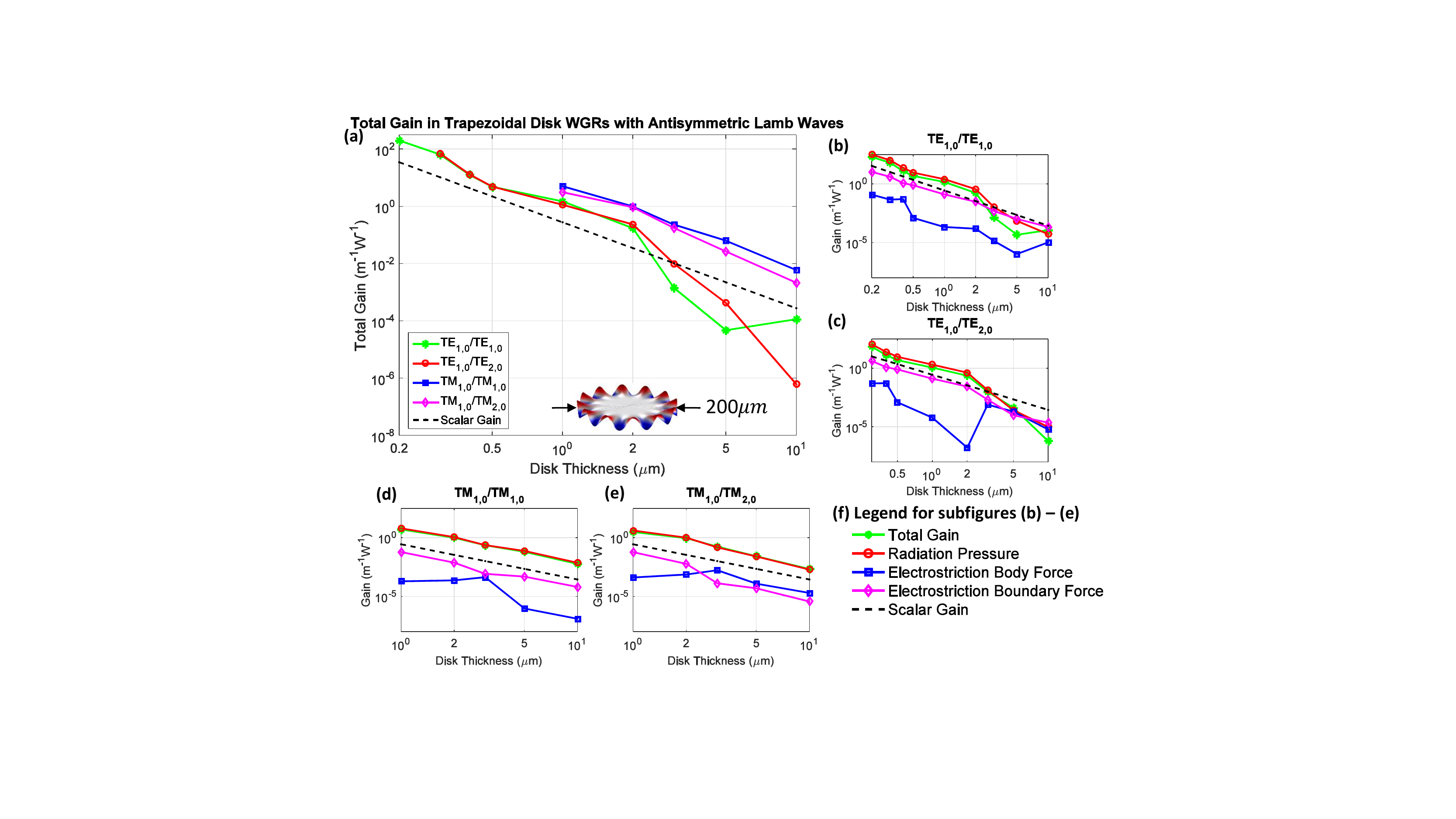}
	\caption{\textbf{SBS in trapezoidal disk WGRs with anti-symmetric Lamb waves ($M=12$) exhibits extremely large gain, and some enhancement over scalar theory}. Significant reduction of the electrostriction body force is observed for TE modes in the $1\,-\,5$ {\textmu}m thickness range due to a transition in the optical mode shape (see Fig.~\ref{fig:8} and main text). (a) The total gain including all forces. (b)-(e) Gain contributions of individual optical forces if they were considered independently. (f) Legend for subfigures (b)-(e).}
	\vspace{-0.15in}
	\label{fig:10}
\end{figure*}

\begin{figure*}
	\centering
	\includegraphics[width=1\textwidth,clip=true,trim=3.3in 1.1in 3in 1in]{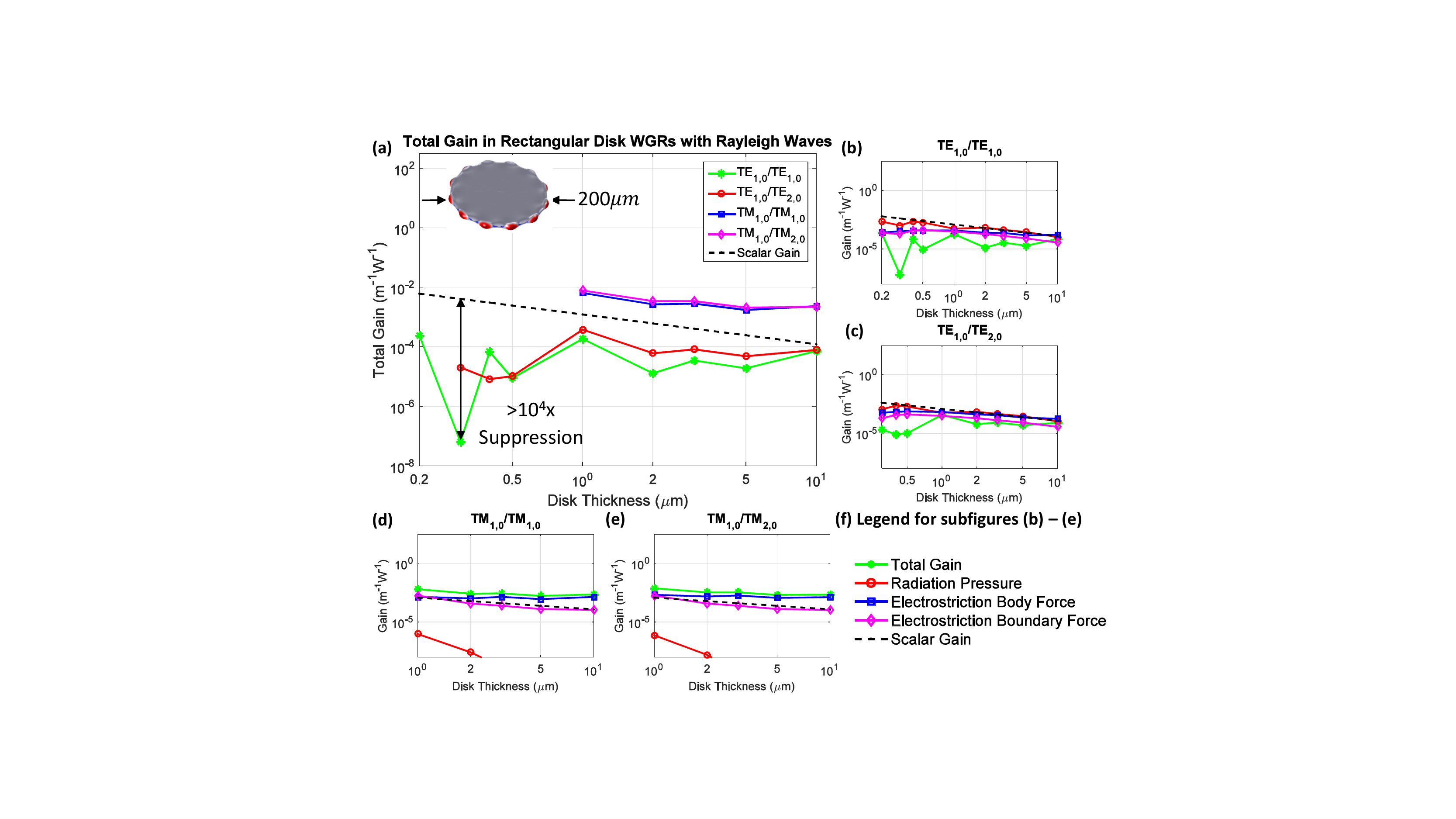}
	\caption{\textbf{SBS in rectangular disk WGRs with Rayleigh waves ($M=12$) exhibits significant gain suppression in TE modes due to cancellation of radiation pressure and electrostriction forces}. (a) The total gain including all forces. (b)-(e) Gain contributions of individual optical forces if they were considered independently. (f) Legend for subfigures (b)-(e).}
	\label{fig:11}
\end{figure*}

Symmetric Lamb modes exhibit periodic compression and rarefaction, much like a longitudinal wave in a bulk material, and intuitively ought to exhibit high SBS gain. However, the regions of axial compression in these acoustic WGMs are accompanied by radial expansion of comparable magnitude. As a result, self-cancellation can occur in the overlap integral of the compressive-only electrostriction force with this mode shape. Similarly, the boundary forces drive the acoustic mode axially via the upper and lower faces, but oppose the mode radially along the outer edge. Rayleigh and anti-symmetric Lamb waves are not affected in this way as they primarily involve displacement along a single direction. Thus, Brillouin gain in disk WGRs with symmetric Lamb waves is not significantly enhanced by thinning the disk.

Another notable trend observable in Fig.~\ref{fig:9} is the decreasing scalar gain as a function of decreasing disk thickness. Symmetric Lamb waves have higher acoustic frequency on thinner substrates, which in this case should decrease the gain quadratically according to Eqs.~\eqref{eq:8}~and~\eqref{eq:13}.

\textbf{Anti-symmetric Lamb wave WGMs}: These modes are characterized by simultaneous upwards or downwards deflections of the upper and lower resonator surfaces \cite{Auld1990}. The results of scalar and full-vectorial computation are presented in Fig.~\ref{fig:10}. We note that very large SBS gain is attainable in thin disks, and is relatively well predicted by scalar theory. Here, the gain increase primarily occurs due to the decreasing acoustic frequency of the anti-symmetric Lamb wave; SBS gain in Eq.~\eqref{eq:8} increases quadratically as acoustic frequency $\Omega$ decreases. This trend matches the known acoustic dispersion for anti-symmetric Lamb waves on thin substrates \cite{Auld1990}.\par

Additionally, we observe significant reduction of the electrostriction body force contributions for thin disks, in the regime where the thickness is comparable to the optical wavelength (Fig.~\ref{fig:10}(b)-(e)). This effect is caused by the disk edge, where the optical electric field orientation transitions from tip-oriented ($>5$ {\textmu}m) to disk-oriented ($<1$ {\textmu}m) as simulated in Fig.~\ref{fig:8}. In this transition region the optical modes change shape such that the electrostriction body force is not able to drive the acoustic mode efficiently.

\subsection*{Brillouin Gain in Silica Rectangular Cross-Section Microdisks}

We now examine Brillouin gain in microdisk resonators of rectangular cross-section, as illustrated in Fig.~\ref{fig:3}. In this case we only analyze Rayleigh and symmetric Lamb-wave acoustic modes. Here, the disk thickness is varied between $0.2\,-\,10$ {\textmu}m using a fixed radius of $100$ {\textmu}m. Overall, we see that the gain predictions for rectangular disks are similar to their trapezoidal counterparts with some deviations induced by the absence of the angled outer edge.\par
\textbf{Rayleigh wave WGMs}: The results of scalar and full-vectorial computation are presented in Fig.~\ref{fig:11}. We note that the gain for axially polarized (TM) mode pairs is similar to the computed values from the trapezoidal disks, i.e. lies in the $10^{-3}\,-\,10^{-2}$ m\textsuperscript{-1}W\textsuperscript{-1} range. In contrast to trapezoidal disks, however, here the radiation pressure contribution to the TM cases is negligible due to absence of the angled outer edge.\par

\begin{figure*}
	\centering
	\includegraphics[width=1\textwidth,clip=true,trim=3.3in 1.3in 3.2in .8in]{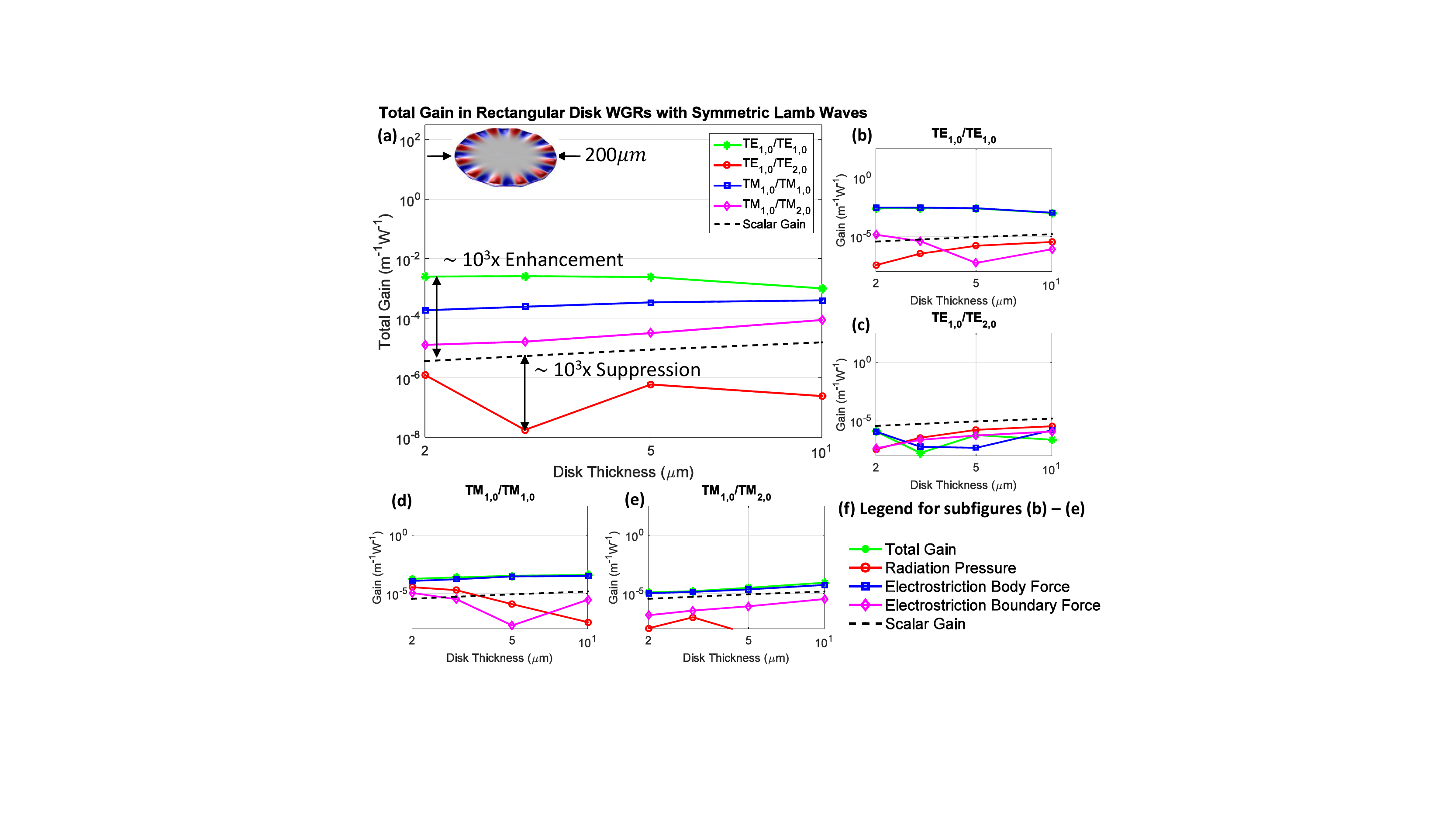}
	\caption{\textbf{SBS in rectangular disk WGRs with symmetric Lamb waves ($M=12$) exhibits low total gain}. Significant suppression of SBS with the higher order TE mode pair is also noted. Acoustic dispersion of these symmetric Lamb modes is responsible the downward trend of scalar gain for thinner disks.  (a) The total gain including all forces. (b)-(e) Gain contributions of individual optical forces if they were considered independently. (f) Legend for subfigures (b)-(e).}
	\label{fig:12}
\end{figure*}

Radially polarized mode (TE) cases do show much higher radiation pressure, on par with electrostriction forces, however the total gain is seen to be much lower than the TM case. This is because in the TE case these forces oppose each other resulting in cancellation of contributions. As a result of this, significant gain suppression($10^4\times$) is expected for the radially polarized inter-modal pair (TE$_{1,0}$/TE$_{2,0}$).

\textbf{Symmetric Lamb wave WGMs}: The results of scalar and full-vectorial computation are presented in Fig.~\ref{fig:12}. We predict the gain to be similar to that in trapezoidal disk WGRs. Again, the gain decreases in thinner disks due to the acoustic dispersion.

\textbf{Anti-Symmetric Lamb wave WGMs}: In a thin rectangular cross-section disk, the optical modes are mirror-symmetric relative to the plane cutting the middle of the disk. The overlap integral (Eq.~\eqref{eq:8}) of the resulting optical forces with anti-symmetric Lamb wave modes yields zero. Thus, we do not analyze this situation in this study as the computational results appear simply as noise of numerical calculation. In stark contrast, as seen above, the Brillouin gain is extremely high in trapezoidal cross-section disks since this mirror-symmetry is broken.

\section{Conclusions}

In this study a full vectorial model for computing the SBS gain associated with surface confined acoustic and optical modes in WGRs is developed. The model is applied to a variety of common WGRs of spherical, shell, trapezoidal disk, and rectangular disk geometries. The overall SBS gain is evaluated by analysis of individual contributions from bulk (electrostriction) and boundary (electrostriction and radiation pressure) optical forces. Using this model, we support our hypothesis that giant enhancements of SBS gain generally take place where optical and acoustic fields are simultaneously confined at a free surface or material interface, and do not require nanoscale features.

Our computational results show that significant boundary forces are present in microscale WGRs, causing both amplification and suppression of the SBS gain of the system. This is contrasted against linear waveguides where nanoscale dimensions \cite{Qiu2013,Shin2013,Rakich2010,Rakich2012} or slotted geometries \cite{VanLaer2014} are essential to achieve similar effects. As expected from previous studies, decreasing the resonator dimensions generally increases the boundary forces resulting in overall gain enhancement. We further note that specific WGR modal configurations (mode shape and polarization) and geometries can be used to enhance or suppress the SBS gain by very large factors, even without considering resonance effects. In particular, we predict extremely high gain enhancement (by a factor of $10^4$) in shell WGRs, and gain suppression (by a factor of $10^4$) in rectangular disk WGRs, for particular sets of modes. It is notable that extreme SBS gain up to $10^2$ m\textsuperscript{-1}W\textsuperscript{-1} can appear in trapezoidal disks without relying on any resonant enhancement effect. We expect these effects to be ubiquitous across WGRs of any dielectric since the enhancement due to boundary forces is a geometric effect. The specific ratio of contribution from boundary forces and bulk electrostriction, however, will be modified by parameters of the particular material, including permittivity, photoelastic coefficients, and acoustic parameters.

Finally, we remind the reader that a further amplification factor of optical-finesse-squared, once for the pump power enhancement and once for the Stokes path length enhancement, applies to all the gains that we compute in this study (see Section~\ref{sec:difference_waveguides_wgrs}). This factor applies only to resonant cavities (including linear resonators) and can approach $10^5\times10^5$ in ultra-high-Q microresonators. Extremely high SBS gain of $4\times10^6$ m\textsuperscript{-1}W\textsuperscript{-1} was recently experimentally measured in a microsphere WGR \cite{Kim15} having $10^5$ finesse factor for both optical modes, suggesting a `bare' gain of $4\times10^{-4}$ m\textsuperscript{-1}W\textsuperscript{-1}. This number is consistent with the enhanced gain (Fig.~\ref{fig:4}) presented in this work and is one order-of-magnitude higher than the prediction of scalar theory. Extreme Brillouin gains approaching $10^{12}$ m\textsuperscript{-1}W\textsuperscript{-1} may thus be reachable in ultra-high-Q trapezoidal microdisk WGRs (Fig.~\ref{fig:10}), which is $10^8$ times greater than the highest predicted \cite{Rakich2012,VanLaer2014} and observed \cite{VanLaer2015} gains in linear waveguides.

\section*{Acknowledgements}
	The authors would like to thank Kewen Han, Junhwan Kim, and Kaiyuan Zhu for helpful discussions towards developing the computational models presented here. We also thank Dr. Peter Rakich for helpful discussion concerning the vectorial theory of SBS gain. This work was funded by a University of Illinois Startup Grant, and in part by the US National Science Foundation (ECCS-1408539) and the US Air Force Office for Scientific Research (FA-9550-14-1-0217).


\bibliographystyle{lpr}
\bibliography{biblio}

\end{document}